\documentclass[a4paper,11pt]{article}
\pdfoutput=1 

\usepackage{jcappub} 

\usepackage[T1]{fontenc} 

\usepackage{amsmath,aas_macros}

\title{\boldmath The haloes that reionized the Universe}

 \author[1]{Nachiket Joshi} 
 \author{\&}

 \author[1]{Mahavir Sharma}
 \affiliation[1]{Department of Physics, Indian Institute of Technology (IIT) Bhilai, 491002, India}

\emailAdd{mahavir@iitbhilai.ac.in}

\abstract{We study the reionization of the Universe due to haloes that host galaxies undergoing bursts of star formation. By comparing the recent results from the James Webb Space Telescope (JWST) with the cosmological hydrodynamical simulation {\sc eagle} at $z\ge 6$, we find that bursty galaxies have specific star formation rate, sSFR~$>10^{-2}$~Myr$^{-1}$, and magnitude, $M_{\rm UV}\leq -17$. Most of them reside in haloes of mass $\sim 10^9$~M$_\odot$ and some in more massive haloes. We then construct the models of escape fraction and find that a skewed Gaussian function with a flat tail towards the high mass end best describes the mean dependence of escape fraction on halo mass, considering the haloes hosting bursty galaxies as the primary drivers of reionization. We implement the models of escape fraction in the code {\sc 21cmfast} to study the progress of reionization and derive the evolution of the mean ionized fraction that agrees well with observations. We also calculate the brightness temperature, spin temperature, and kinetic temperature and further study the spatial fluctuations in these quantities to gain insights into the progress of reionization. We compute the 21 cm power spectrum and predict a peak in power at $180$~MHz corresponding to redshift, $z\approx6.8$, that is testable by the upcoming Square Kilometre Array (SKA). Our findings suggest that the Universe was reionized by the haloes of $\gtrsim 10^{9}$~M$_\odot$.
}

\begin{document}
\maketitle
\flushbottom

\section{Introduction}
The epoch of reionization (EoR) is a period in the history of the Universe when the first stars and galaxies emerged and provided the ionizing photons to reionize the neutral intergalactic medium (IGM) \cite{Naoz2006, Loeb2001}. Earlier studies based on the absorption troughs in quasar spectra indicated that reionization occurred at $z>6$ \citep{Rauch1998, Fan2003, Fan2006}. The exact timing of reionization was further narrowed down with the space-based measurements of the electron-scattering optical depth by the {\it Planck} satellite. The latest measurements from {\it Planck} indicate that the Universe was 50\% ionized at redshift, $z = 7.64$
\cite{Aghanim2020}. Apart from the timing of reionization, identifying the sources that reionized the Universe is a major focus of research \cite{Barkana2016, Sharma2016, Sharma2017, Sokasian2004, Haardt2012, Robertson2013, Madau2015, Mitra2018}.
Various studies indicate that the galaxies within the first haloes led the process of reionization \cite{ Robertson_2015, Bouwens_2015, Bromm2009, Barkana2001}. 
However, haloes have a range of mass, and galaxies within them also exhibit a huge diversity \cite{Shibuya2015, Furlong2015}. 

The properties of galaxies that reionized the Universe and their host haloes are under investigation \cite{Finkelstein2015, cohen17,endsley2023, Nakane2023, Simmonds2024, Atek2015, Bouwens2017, Atek2024}.
Recent theoretical work \citep{Sharma2016, Sharma2017, Sharma2018, Qin2023}, and the latest observation by the James Webb Space Telescope (JWST) has made significant progress in revealing the galaxies that reionized the Universe \citep{Looser2023, Simmonds2024, Nakane2023, Enders2023}. However,  there is still ambiguity about the dark matter haloes that host such galaxies. Galaxies develop within the potential wells of dark matter haloes via gas cooling that leads to star formation \citep{Bromm2009, Barkana2001, Freese2009}. The ionizing photons that drive reionization are produced by the massive stars formed during the initial bursts of star formation \citep{Barkana2001, Sharma2016, Sharma2017}. 

The models of reionization depend on a critical parameter known as the escape fraction \cite{Gnedin_2008, Simmonds2024, Sharma2016, Calabro2024}, which connects the number of ionizing photons produced by massive stars in a galaxy to the number that actually leak out of the galaxy \citep{Gnedin_2008, Ocvirk2016, Lewis2020, Mesinger2011, Ferrara2013, Wise_2009, Kostyuk2023}. The galaxies are complex systems. They have a reservoir of gas that fuels star formation and depletes as a result but then is replenished through cosmological infall \cite{Lacey1985, Matteucci1989, White1991}. A disc develops after a sufficiently long period, as is evident for galaxies at low redshifts. Almost all the photons are absorbed in such an optically thick system, indicating a negligible escape fraction that agrees with most of the attempted direct measurements of escape fraction, which have yielded null results \cite{Japelj2017, Leitherer1996, Heckman2011, Shapley2006, Siana2007, Vanzella_2010, Nestor_2011, Boutsia_2011}. However, recent detection of high escape fraction in the local Universe  \cite{Izotov2016, Enders2023, Melinder2023} hints that the escape fraction is tied unsurprisingly with the nature of star formation in a galaxy \citep{Sharma2016, Sharma2017, Calabro2024}.

The escape fraction is linked with the intensity of star formation and subsequent feedback by stars and supernovae \cite{Fernandez2011, Trebitsch2017, Kimm2014}. Also, the pictures of galaxies having homogeneous discs \citep{Gnedin_2008} with uniform densities may be too simplistic, specifically for the galaxies at high redshift \citep{Wise2014, Yajima2011, Simmonds2024}. 
The cosmological simulations of galaxy formation \citep[e.g.][]{Schaye2015}, the large-scale surveys such as the SDSS \cite{Shibuya2015}, and more recent breakthrough observations by the James Webb Space Telescope (JWST) \citep{Robertson2023a,robertson2023b, Simmonds2024}; all indicate that early galaxies were much more violent, with a significantly bursty star formation. The closest counterparts of high redshift galaxies are likely the extremely active nearby starbursts undergoing violent episodes of star formation \cite{McQuinn2023, Harikane_2023, Bouwens2023}. The extreme starburst activity is an exception at low redshift, whereas it is a norm at high redshift prior to reionization \citep{Shibuya2015, Sharma2016, Sharma2017, Faucher-Giguere2018,  Looser2023, Calabro2024}. 
A burst of star formation, specifically in a compact galaxy at high redshift, disrupts the distribution of gas and opens up the belly of the galaxy through which almost all of the ionizing photons can leak out, and the escape fraction at that moment can easily be more than 10\% \citep{Borthakur2014, Izotov2016, Schaerer2019}.

Theoretical studies of reionization often deployed a constant escape fraction for all the galaxies \citep[e.g.][]{Mesinger2011}, or more recently, an escape fraction that decreases with increasing halo mass \citep{Park2019}. Naturally,  the majority of the reionizing photons then would be from the low mass `faint' galaxies, leading to the traditional viewpoint that faint galaxies reionized the Universe \citep{Finkelstein2019, Atek2024, Finkelstein2015, Atek2015, Bouwens2017}. However, recent studies challenged this viewpoint through attempts at more astrophysics-oriented modelling of escape fraction by including the consequences of bursty behaviour of high redshift galaxies \citep{Sharma2016, Sharma2017}. These studies modelled the escape fraction of pre-reionization galaxies by utilising the star formation rate surface density as a criterion for bursts that puncture the gas distribution in galaxies. The outcome was intriguing, that the majority of the ionizing photons were produced by the relatively brighter galaxies rather than the faintest ones \citep{Sharma2016, Robertson2023a, Looser2023, Calabro2024}. 

For the galaxies at low redshifts, it is known that statistically, high brightness or SFR indicates a galaxy with a high stellar mass  \citep{Marinacci2013, Speagle2014, Popesso2023}; this is how, after all, the main sequence of galaxies is defined \citep{Marinacci2013, Speagle2014, Popesso2023, Tacchella2020}. Further, sufficient time may be needed for the emergence of a main sequence, though its existence is not surprising as it is a relation between the SFR and its time integral, the stellar mass \citep{Tacchella2020, Popesso2023}. A further correlation of SFR (or stellar mass) with halo mass may exist, but its extrapolation to high redshift may not be valid. The scatter in the stellar to halo mass relation, in fact, increases with redshift \citep[e.g.][]{Shuntov2022,Golden2022}, and it also increases towards the low halo mass end \citep{Matthee2017}. Therefore, not only do the low-mass haloes host faint galaxies, but they also host relatively brighter galaxies with magnitude, $\rm M_{\rm UV} \le -17$, particularly at high redshifts, as indicated by the recent detections of many low-mass yet brighter galaxies by the JWST \citep{Robertson2023a, endsley2023, Simmonds2024, Looser2023}. 
The instantaneous SFR in low-mass haloes is not monotonous, but it fluctuates with intense bursts followed by a quiescent period of almost zero SFR \citep{Kimm2017, Trebitsch2017, Tachella16, Faucher-Giguere2018, Tacchella2020, Sharma2018CEMP, Sharma2019, Sun2023, Nikolic2024}. 
Therefore, it is unlikely that the early galaxies would precisely follow the expected Halo mass, stellar mass, and SFR correlations. Even if they show signs of such correlations, still the scatter would be the prominent feature. This crucial point is often missed in the semi-analytical models \citep{Mesinger2011, Park2019} since one has to rely on the correlation between the halo mass, stellar mass, and the SFR to estimate the photon budget. 

In this paper, we first focus on identifying the haloes that host bursty galaxies by studying the properties of galaxies detected recently by the JWST and comparing them with the early galaxies and haloes in the cosmological hydrodynamical simulation {\sc eagle} \citep{Schaye2015}. We then use our findings to model the progress of reionization using the semi-numeric code {\sc 21cmfast} \citep{Mesinger2011, Park2019}. We study the progress of reionization if the Universe is reionized by the haloes that host the bursty galaxies, and we investigate in-depth the corresponding evolutions of mean neutral hydrogen, mean brightness temperature, the patchiness of these quantities, and the power spectrum at the redshifted 21 cm wavelength. Our main objective is to find the haloes that reionized the Universe (HRUs). 

The paper is organized as follows. Section 2 presents our findings on the star formation in early galaxies and their host haloes based on the recent results from the JWST and cosmological simulation {\sc eagle}. Section 3 describes our models that utilize specific haloes as drivers of reionization and presents our main results from {\sc 21cmfast} in subsequent subsections. We discuss and summarize our findings in section 4.

\section{Star formation in pre-reionization haloes}
\subsection{Scatter in the SFR and $M_{\rm UV}$ at the low mass end: dwarfs aren't always faint}

The recent observations by the JWST Advanced Deep Extragalactic Survey (JADES) report galaxies with signatures of bursty star formation \citep{Looser2023}. The JADES galaxies have UV magnitudes $M_{\rm UV}$  roughly between $-17$ and $-21$, and a majority of them have low stellar masses in the range $10^6$ to $10^8$~M$_\odot$. Another JWST observational survey, UNCOVER, reports `faint' low-mass galaxies with $M_{\rm UV}\geq -17$ and claims such galaxies provided sufficient ionizing emissivity to reionize the Universe \citep{Atek2024}. The two sets of observations differ in the brightness of galaxies despite both corresponding to the low mass end. 

\begin{figure}
    \centering
    \includegraphics[scale = 0.7]{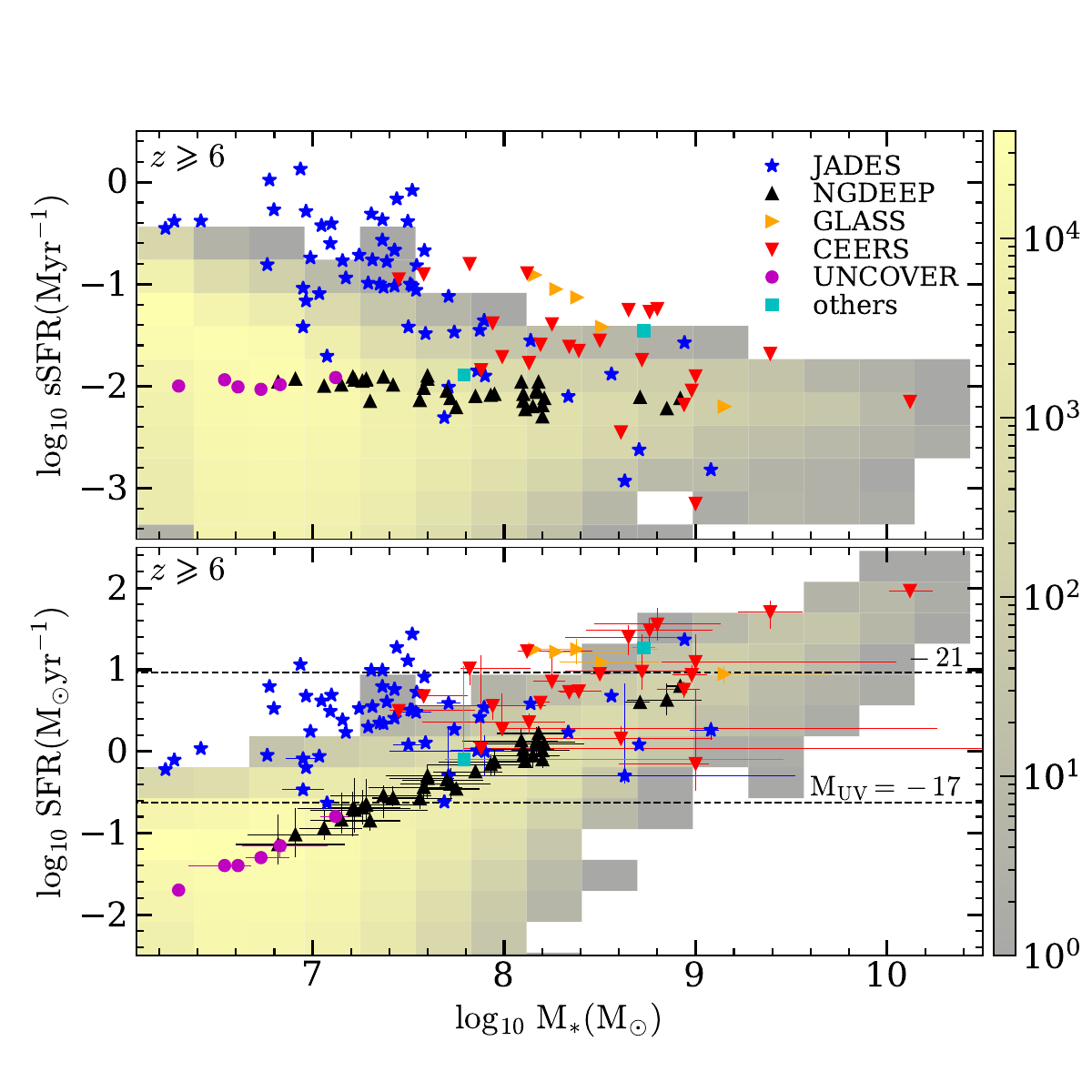}
    \caption{The star formation rate, SFR (bottom panel),  and the specific star formation rate, sSFR = SFR/$M_{\rm \star}$ (top panel), as a function of the stellar mass, $M_{\rm \star}$, for the galaxies at, $z\ge 6$, observed by various JWST programs (blue stars for JADES \citep{Robertson2023a, Looser2023, Bunker2023, Jones2023, Curti2023}, black triangles for NGDEEP \citep{Morales2023}, orange right pointing  triangles for GLASS \citep{Nakajima2023}, right down pointing triangles for CEERS \citep{Nakane2023, Nakajima2023, Jung2023, Haro2023, Harikane2024, Zitrin2015, Fujimoto2023, Tang2023, Larson2023, Sanders2023}, purple circles for UNCOVER \citep{Atek2024}, and cyan squares \citep{Bunker2023b, Harikane2024, Haro2023}), compared with the  galaxies in the {\sc eagle} simulation  shown as a 2D histogram in the background for the distribution of galaxies \citep{Schaye2015}. The scatter increases towards the low mass end leading to a degeneracy in the SFR and UV magnitudes at the low mass end. Therefore, low mass end has the expected faintest galaxies (purple circles)\citep{Atek2024}, however it also has relatively brighter galaxies (blue stars) with a above-normal sSFR due to burstiness (see top panel) \citep{Looser2023, Robertson2023a}.}
    \label{fig_first}
\end{figure}

Figure~\ref{fig_first} lower panel shows the SFR for the galaxies detected by various JWST programs \citep{Robertson2023a, Bunker2023, Jones2023, Curti2023, Looser2023, Nakane2023, Morales2023, Atek2024, Nakajima2023, Jung2023, Haro2023, Harikane2024, Zitrin2015, Fujimoto2023, Tang2023, Larson2023, Sanders2023, Bunker2023b, Haro2023}. For comparison, the SFRs for the galaxies in the cosmological simulation {\sc eagle} are shown as a 2D histogram in the background. The most massive galaxies with stellar masses above $10^{8}$~M$_\odot$ have highest SFRs ($\ge 10$~M$_\odot$yr$^{-1}$), and UV magnitudes $\le-21$. On the other hand, at the low-mass end, there is a degeneracy in the SFR (and UV magnitudes). The data points corresponding to galaxies from NGDEEP \citep{Morales2023}, and the faint galaxies detected by \citep{Atek2024} follow an SFR-M$_\star$ correlation indicating a main sequence. However, the galaxies from JADES \citep{Robertson2023a, Looser2023, Bunker2023, Jones2023, Curti2023} exhibit higher SFRs. Most of the JADES galaxies are brighter with $M_{\rm UV}$ in the range $-17$ to $-21$. These galaxies are undergoing bursts as claimed by \citep{Looser2023}, which is the reason for their relatively high brightness despite most of them having low stellar mass, same as \citep{Atek2024}. 

The picture is further clear in the upper panel of Fig.~\ref{fig_first}. It shows the specific star formation rate, sSFR, for the JWST galaxies compared with the {\sc eagle} simulation. The horizontal main sequence is apparent both in the observations and {\sc eagle}. The galaxies from UNCOVER \citep{Atek2024} follow the main sequence, and they lie at the faint low-mass tail. However, there is a considerable scatter at the low-mass end in the data, as the bursty galaxies from JADES occupy the upper left quadrant with their above-normal sSFR $> 10^{-2}$~Myr$^{-1}$. Some galaxies from CEERS \citep{Nakane2023, Nakajima2023, Jung2023, Haro2023, Harikane2024, Zitrin2015, Fujimoto2023, Tang2023, Larson2023, Sanders2023} and GLASS \citep{Nakajima2023} with a stellar mass between $10^7$ and $10^9$~M$_\odot$, and $M_{\rm UV}< -21$, also exhibit above-normal sSFRs.

Thus, it is evident that the low-mass galaxies are not always faint. During bursty episodes, they are brighter. The star formation begins in low mass haloes, and initially, the SFR fluctuates between peaks and troughs \citep{Faucher-Giguere2018, Tachella16, Sharma2019}.  The Universe was likely reionized by the bursty brighter galaxies with sSFR $> 10^{-2}$~Myr$^{-1}$ and $M_{\rm UV} \le -17$; already being detected by the JWST \citep{Robertson2023a, Looser2023}. Indeed, in an earlier study \citep{Sharma2016}, we predicted that almost all ($\approx 90\%$) ionizing photons required for reionization were provided by the `brighter' galaxies with magnitudes $M_{\rm UV}< -16$, and  JWST would be able to detect them.

\begin{figure}
    \centering
    \includegraphics[scale = 0.7]{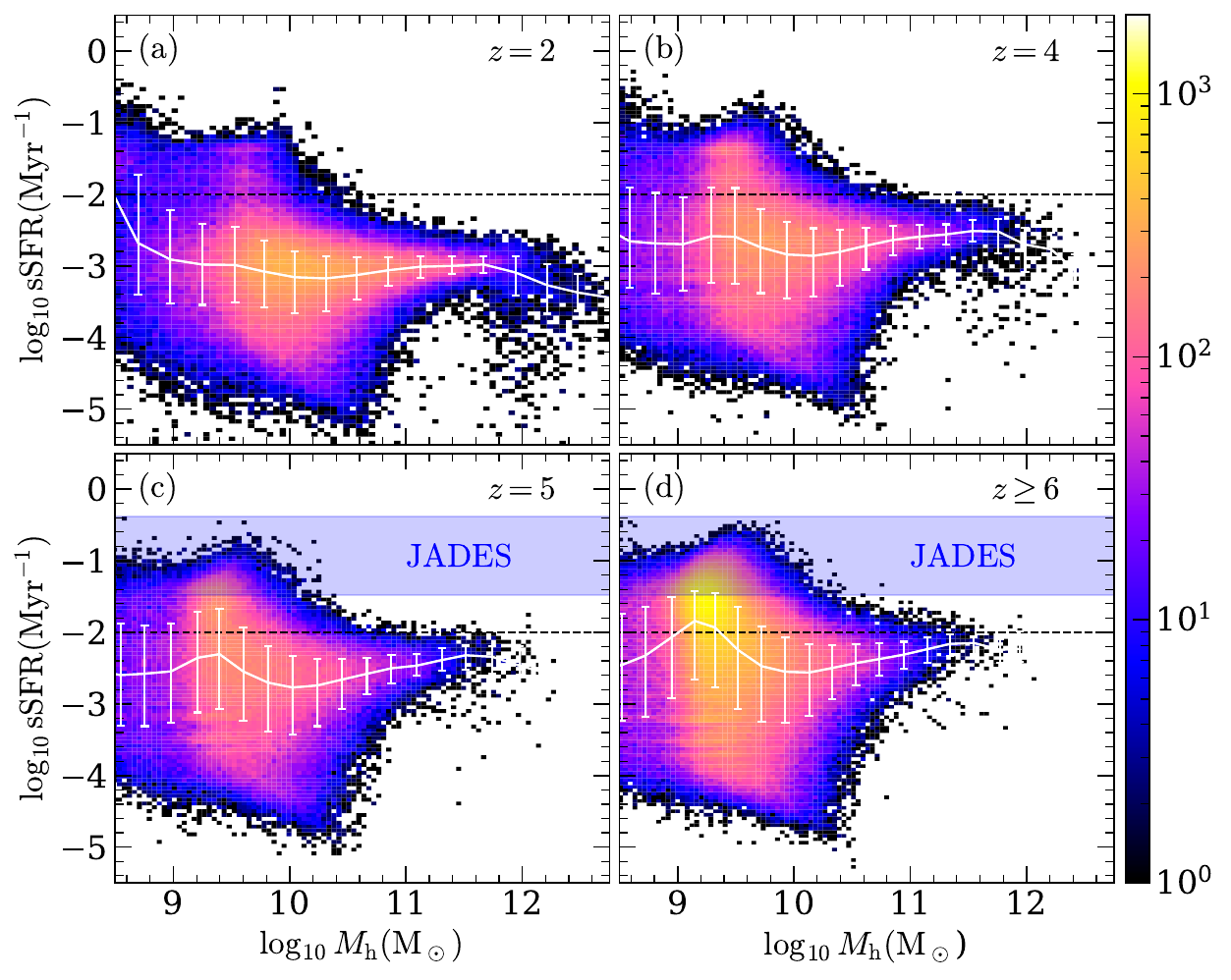}
    \caption{The specific star formation rate, sSFR = SFR/$M_{\rm \star}$, as a function of the halo mass, $M_{\rm h}$, for the galaxies in the {\sc eagle} simulation \citep{Schaye2015}, at redshift $z=2$ (panel a), $z=4$ (panel b), $z=5$ (panel c) and $z\ge 6$ (panel d). The white curve shows the median sSFR in a bin of $M_{\rm h}$ with the error bars showing the range of 20th to 80th percentile. At lower redshifts (upper panels), the sSFR for all the halo masses is scattered around  a horizontal ridge (i.e. main-sequence) at ${\rm sSFR}=10^{-3}$~Myr$^{-1}$. At high redshift ($z>6$), the main sequence shifts up overall and there is a peak in the median sSFR at a halo mass of approximately $10^{9}$~M$_\odot$. For comparison, the recently detected galaxies by JWST Advanced Deep Sky Survey (JADES) are shown as a horizontal blue band covering the range from 20th to 80th percentile of their sSFR distribution \citep{Robertson2023a, robertson2023b, Looser2023, Harikane_2023}.}
    \label{fig_SFR_Mh}
\end{figure}

\subsection{The halo mass corresponding to the bursty galaxies with highest sSFR}

Determination of halo mass observationally for high-$z$ galaxies is challenging. The existing stellar-to-halo mass relation is also uncertain at high redshift. 
The cosmological N-body simulations of galaxy formation are a valuable tool to relate the properties of galaxies with dark matter haloes, as the dark matter is the basic fabric of these simulations \citep{Springel2005, Vogelsberger2014, Schaye2015}. The {\sc eagle}  \citep{Schaye2015} simulation models the formation of galaxies within dark matter haloes and reproduces their properties across cosmic time. {\sc eagle} is based on a modified version of the smooth particle hydrodynamic (SPH) code {\sc gadget 3} with improvements in time-stepping and subgrid physics \citep{Schaller2015, Crain2015}. The feedback is implemented through an updated energy-based scheme that depends on gas density and metallicity. The usual inefficiency due to resolution in SPH schemes is compensated for by an additional injection of energy at the SPH particle level \citep{DallaVecchia2008, Schaye2010, Crain2015}. The star formation is incorporated using a version of the Kennicut-Schmidt law \citep{Kennicutt1998}, in which the SFR for a star particle is determined from a relation between the gas surface density and the pressure governed by an equation of state \citep{Schaye2008}. Accordingly, the gas particle is converted to a collisionless star particle stochastically in time. Herein we use the {\sc eagle} run L100N1504 with dark matter particle mass $9.7\times 10^6$~M$_\odot$, and the gas-particle mass $1.8\times 10^6$~M$_\odot$. The simulation reproduces the cosmic star formation history, the stellar-to-halo mass relation, and the physical properties of galaxies in detail \citep{Schaye2015}, and it serves as a benchmark for galaxy formation and evolution.

Figure~\ref{fig_SFR_Mh} shows the specific star formation rate, sSFR, plotted with halo mass, $M_{\rm h}$, for the galaxies in the {\sc eagle} simulation. At low redshift (panel a), the main feature is that most of the galaxies lie around a horizontal ridge, known as the main sequence of galaxies in the literature \citep{Marinacci2013}. However, there is a striking difference at high redshift ($z\ge 6$),  leading to a higher scatter and a peak in median sSFR towards the low mass end (panels c and d). Therefore, at high redshift, a significant number of galaxies exhibit extreme sSFRs ($\gg 10^{-2}$~Myr$^{-1}$) both in {\sc eagle} and in the real Universe (Fig.~\ref{fig_SFR_Mh} and \ref{fig_first}), and the distribution of $M_{\rm h}$  peaks at $\sim 10^9$~M$_\odot$. The same set of galaxies in Fig.~\ref{fig_first} corresponds to a slightly wider range of stellar masses, $10^6$ to $10^8$~M$_\odot$, as the stellar mass is sensitive to the nature of star formation.

The above normal sSFR at the low mass end reflects the violent bursty star formation. The bursts likely occur when an adequate reservoir of gas builds up. For infant galaxies, the threshold reservoir of gas for initial bursts is likely to be achieved for a specific halo mass since, before the beginning of star formation, the net baryonic accretion is tightly correlated with the dark matter accretion. The initial bursts are unique as the stellar mass is negligible; therefore, the values of sSFR are very high. 

For the episodic star formation at the beginning in a newly formed galaxy \citep{Sharma2016, Sharma2018CEMP, Faucher-Giguere2018}, an analogy can be drawn with an engine. In the beginning, the engine cranks and has episodic breathing and puffing behaviour, and subsequently, it settles into a smooth operation. Similarly, the galaxies are initially compact and dense, waiting for star formation to ensue. The initial burst, as it occurs, is extreme for a small galaxy, and they can clear almost all of the gas, causing a quenching of the SFR. A peak in SFR followed by a rapid decay occurs repeatedly but with less and less intensity in sSFR as time progresses \citep{Sharma2019, Tacchella2020, Looser2023}. Eventually, the SFR settles into a monotonous evolution as the galaxy grows in age, mass, and size \citep{Tacchella2020}.

Simulations such as {\sc eagle} successfully reproduce galaxies' properties in detail. However, they may have limitations due to the resolution. The resolution may influence features such as the scatter and the peak in median sSFR at the low mass end. The peak in the sSFR in Fig.~\ref{fig_SFR_Mh} occurs roughly at a halo mass of $2\times 10^9$~M$_\odot$, that has more than $200$ dark matter particles, though the haloes below $10^9$~M$_\odot$ have fewer than $100$ particles and their properties may be resolution limited. However, a similar large scatter in SFR due to stochastic bursty star formation is reported in earlier high-resolution studies with {\sc eagle} that focused on the origin of metal-poor stars in the low-mass progenitors of the Milky Way \citep{Sharma2018CEMP,Sharma2019}. Furthermore, the bursty behaviour is a universal feature for the newly formed galaxies in many simulations of galaxy formation \citep{Trebitsch2017, Sparre2017, Sun2023, Sharma2018CEMP},  and JWST also finds the same features, indicating them to be a characteristic of the star formation in the first galaxies in the real Universe \citep{Looser2023}.

The bursts, specifically the extreme ones, have significant consequences for reionization. The ionizing photons in galaxies are produced by massive stars that are short-lived ($\sim 10$~Myr). In galaxies like the Milky Way, most ionizing photons currently produced are absorbed within a few parsecs in the dense interstellar medium. The computational studies to model the leakage of photons from discs of galaxies such as the Milky Way yield negligible escape fraction in agreement with the observational findings \citep{Gnedin_2008}. The trend continues even to redshifts up to $3$ where the attempts to measure the Lyman continuum from the Lyman break galaxies have returned negligible escape fractions \citep{Nestor_2011, Shapley2006}. However, recently, Lyman continuum leakage with high escape fractions has been detected in starburst galaxies with high $\Sigma_{\rm SFR}$ and sSFR \citep{Borthakur2014, Izotov2016, Calabro2024}. Although such measurements are still a minority, there is little doubt that galaxies can have porous channels due to violent starbursts through which ionizing photons escape \citep{Wise2014, Sharma2016, Calabro2024}.

To conclude this section, the bursty galaxies have extreme sSFR~$>10^{-2}$~Myr$^{-1}$, and they are typically brighter than  $M_{\rm UV} = -17$ (Fig.~\ref{fig_first}). Comparison of the pre-reionization bursty galaxies recently detected by JADES \citep{Looser2023} with {\sc eagle} (lower panels of Fig.~\ref{fig_SFR_Mh}), shows that the distribution of the halo mass for bursty galaxies peaks at  $\sim 10^9$~M$_\odot$. These galaxies are the prime candidates to be the significant Lyman-continuum leakers \cite{Robertson2023a, Nakane2023}.  
Whether the photons supplied by haloes in a narrow mass range are sufficient to reionize the Universe needs investigation. Moreover, if a subset of halo mass reionized the Universe, there may be intriguing observational consequences, specifically on the 21 cm power spectrum. We study these aspects in the following section.

\section{Reionization of the Universe and the power spectra of neutral hydrogen}

In this section, our objective is to study the progress of reionization driven by specific haloes in a narrow range of halo-mass. 
The main ingredient for reionization is the number of Lyman-continuum photons available in the intergalactic medium (IGM). The photons are generated by massive stars in galaxies that form within the first dark matter haloes \citep{Zaroubi2013}. The number of dark matter haloes of a specific mass (i.e. the halo mass function) can be obtained by the semi-analytic methods such as the extended Press-Schechter formalism \citep{Press1974}, or from the full cosmological simulations such as {\sc millenium} \citep{Springel2005}, {\sc illustris} \citep{Vogelsberger2014} or {\sc eagle} \citep{Schaye2015}. The crucial step is then relating a star formation rate to a halo that ideally reproduces the luminosity functions of high-redshift galaxies from observations. Once an SFR is available corresponding to a halo, one can use a model for the fraction of photons that escape into the IGM to finally solve the radiative transfer problem in the evolving IGM to reionize the Universe and keep it ionized subsequently. 

Various earlier studies have undertaken the task described above \citep{Park2019, Mesinger2007, Mesinger2011, Mutch2016, Garaldi2022}, with each method having its merit or demerit. In this paper, to study reionization and to get the statistical information of diminishing neutral hydrogen islands, we use the popular semi-numeric code {\sc 21cmfast} \citep{Mesinger2011, Park2019}, that solves the radiative transfer problem in an evolving Universe. It first generates the statistical set of haloes using the halo mass function, an outcome of the initial density perturbation field having a Harrisson-Zeldovich power spectrum \citep{Harrison1970, Zeldovich1972}. Each halo is then assigned a stellar mass and an SFR as per a physical model with constraints such as the observed luminosity functions of galaxies and the observed evolution of the cosmic SFR. Finally, for each halo with a given SFR, a constant number that gives the photons per stellar baryon is used. The photon production from each halo is further multiplied by the crucial parameter of escape fraction, $f_{\rm esc}$. The earlier {\sc 21cmfast} studies considered it be constant \citep{Mesinger2011} but later studies also experimented with a halo mass dependent escape fraction \citep{Park2019, Greig2015}. 
In this paper, for our {\sc 21cmfast} runs, we use the default parameters as in the fiducial model by \citep{Park2019}; however, we implement a new halo mass dependent escape fraction, $f_{\rm esc}$, as described in the following section.

\subsection{Models of escape fraction as a function of halo mass}
\begin{figure}
   \centering
   \includegraphics[scale = 0.7]{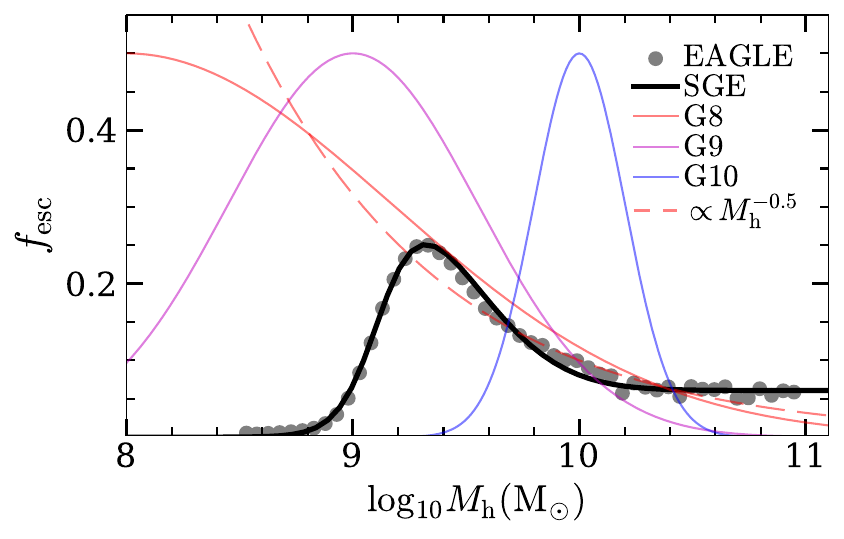}
    \caption{The escape fraction, $f_{\rm esc}$, as a function of the halo mass, $M_{\rm h}$, for the models considered in this work. The gray-filled circles represent the mean escape fraction calculated for the galaxies in the {\sc eagle} simulation to account for the contribution only from the highly bursty galaxies. The solid black curve is a fit to the gray circles (Eq.~\ref{eq_SG}), and we name this model SGE. We have three additional models, given by simple Gaussian function that peaks at $10^8$, $10^9$, and $10^{10}$~M$_\odot$, and we name them G8 (red line), G9 (magenta line), and G10 (blue line) respectively. For the Gaussian models, the maximum value of escape fraction, $f_{\rm max}=0.5$, and width is calibrated such that the reionization finishes at redshift $z=6$. For the SGE model, the only free parameter is $f_{\rm max}$, which is set to $0.25$ so that the reionization finishes at $z=6$. For comparison, the model with $f_{\rm esc}\propto M_{\rm h}^{-0.5}$ by \citep{Park2019} is shown as a red dashed line. 
    }
    \label{fig:fescMh}
\end{figure}

In {\sc 21cmfast} \citep{Mesinger2011, Park2019}, the number of ionizing photons per baryon produced inside a region is given by,
\begin{equation}
    n_{\rm ion} = \Bar{\rho_{b}}^{-1}\int_{0}^{\infty} dM_{\rm h} \frac{dn(M_{\rm h},z|R,\delta_{R})}{dM_{\rm h}}f_{\rm duty}M_\star f_{\rm esc}N_{\rm \gamma /b} \,
\end{equation}
where $ \Bar{\rho_{\rm b}}$ is the mean baryon density, $M_\star$ is the stellar mass in a halo of mass $M_{\rm h}$, $f_{\rm esc}$ is the escape fraction of ionizing photons from a halo, $f_{\rm duty}$ is the duty cycle which is the fraction of the haloes that host stars, $N_{\rm \gamma /b}$ is the number of ionizing photons per stellar baryon which in {\sc 21cmfast} is set to a fiducial value of 5000 \citep{Park2019}.

The photon budget is crucially dependent on the astrophysical parameters, $M_{\star}$ and $f_{\rm esc}$, which are further combined into the {\sc 21cmfast} specific definition of the `ionizing efficiency',  
$  \zeta = f_\star f_{\rm esc}N_{\rm \gamma /b} $ \citep{Park2019}, 
where $f_\star =  M_\star/M_{\rm h}$. The earlier studies based on {\sc 21cmfast} used constant values of $f_{\rm esc}$ and $f_\star$, hence a constant $\zeta$ \citep{Greig2015, Greig2017, Greig2018, Mesinger2011}. However, a better model of stellar to halo mass ratio can be used with improved luminosity functions from the Hubble deep field (HDF) and now from JWST. Furthermore, the escape fraction too is not necessarily constant. Recent studies debate whether it decreases or increases with the halo mass \citep{Greig2015, Ma2015, Kostyuk2023}. Nevertheless, it is certain that the escape fraction depends on the halo mass. The power-law slopes of $f_{\rm esc}$ and $f_\star$ should be constrained by observations. Recent {\sc 21cmfast} study \cite{Park2019} indeed implements halo dependent $f_\star$ motivated by the high redshift luminosity functions from HDF \citep{Bouwens2019}, and they also implement a halo dependent $f_{\rm esc}$, such that the $f_\star$ is considered to be increasing with the halo mass with a power-law slope $\alpha$, while the escape fraction is considered to be decreasing with the same slope so that the ionizing efficiency $\zeta$ is effectively constant \citep{Park2019}.

 In this work, we implement, $\zeta = f_{\rm esc} f_\star N_{\rm \gamma/b}$, and,$f_\star = f_{\star,10}[M_{\rm h}/M_{10}]^{\alpha_\star}$, same as in the fiducial {\sc 21cmfast} model of \citep{Park2019} with  $N_{\rm \gamma/b} =5000$, $\alpha_\star = 0.5$, and  $f_{\star,10}=0.05$ is the ratio of stellar to halo mass for $10^{10}$~M$_\odot$ haloes. We also retain $f_{\rm duty}$ with its turnover mass, $M_{\rm turn}$, same as in \citep{Park2019}. However, we implement new models for the escape fraction, $f_{\rm esc}$,  motivated by the bursty nature of galaxies at high redshift.

Recent theoretical work and latest observations on escape fraction \cite{Sharma2016, Sharma2017, Izotov2016, Schaerer2019, Naidu2022}, indicate that $f_{\rm esc}$ is tied to the surface density of star formation, $\dot \Sigma_{\star}$, which is an indicator of outflows that can open-up channels through which photons can escape \citep{Sharma2016, Sharma2017, Calabro2024}. Therefore, $f_{\rm esc}$ is likely high in compact galaxies undergoing violent episodes of star formation, which is a norm at high redshift \citep{Calabro2024}. The expected and observed dependence of the escape fraction on the properties of galaxies such as the $\Sigma_{\rm SFR}$ and $\rm sSFR$ \citep{Calabro2024} can be used to implement the escape fraction in models of reionization \citep{Sharma2016, Sharma2017, Naidu2020}. For example, in a recent compilation by \citep{Calabro2024} indicate that the measured $f_{\rm esc}$ are significant, $\gtrsim 0.1$, for galaxies with sSFR $\gtrsim 10^{-2}$~Myr$^{-1}$.

A positive correlation between sSFR (or $\dot \Sigma_\star$) and halo mass would indicate $f_{\rm esc}$ increasing with halo mass, and a similar trait for $\zeta$, in contrast to \cite{Park2019}, which was utilized in a study by \citep{Sharma2016}. 
However, the smooth trends between SFR and halo mass might not hold for the early infant galaxies, as discussed in the previous section (see Fig.~\ref{fig_SFR_Mh}). Nonetheless, the escape fraction would be high for galaxies undergoing violent bursts of star formation, and Fig.~\ref{fig_SFR_Mh} shows that such galaxies reside in a narrow range of halo mass at the low-mass end.

We construct a model of the escape fraction as a function of the halo mass by using the galaxies in the {\sc eagle} simulation with the following criteria. If a halo of mass $M_{{\rm h},i}$ in a bin of halo mass (in Fig.~\ref{fig_SFR_Mh}) satisfy ${\rm sSFR} > 10^{-2}$~Myr$^{-1}$, we assign an escape fraction value of $f_{{\rm esc},i}=C$, and $0$ otherwise; and then we take a mass weighted mean of escape fractions of all the haloes in the bin of halo mass, to get, $f_{\rm esc} = \sum_i f_{{\rm esc},i}M_{{\rm h},i}/\sum_{i} M_{{\rm h},i}$. $C$ is expected to be negligible in quiescent galaxies and significantly higher in leaky starbursts charaterized through extreme $\Sigma_{\rm SFR}$ or sSFR \citep{Borthakur2014, Izotov2016, Sharma2018, Schaerer2019, Calabro2024, robertson2023b}. We use a constant value for $C$ to represent the bursty galaxies. Even for a constant $C$, the average escape fraction evolves with redshift, as the starbursts are rarer at low redshift but their prevalence increases with redshift.

The escape fraction corresponding to each bin of halo mass thus obtained is shown in Fig.~\ref{fig:fescMh} with gray-filled circles, where the constant $C$ is set such that the peak value of $f_{\rm esc}$ curve as a function of $M_{\rm h}$ is $f_{\rm max}= 0.25$ to finish the reionization at $z=6$ when implemented in {\sc 21cmfast}. We fit the data points with a skewed Gaussian function with a flat tail, written as,
\begin{equation}
    f_{\rm esc} = f_{\rm max}  \left[ \frac{ e^{-((x-\mu)/\sigma)^{2}}}{1+e^{-a(x-\mu)/\sigma}}  + \frac{b}{1+e^{-a(x-\mu)/\sigma}} \right]
    \label{eq_SG}
\end{equation}
where, $x=\log_{10} M_{\rm h}$, $a=5.945$, $b=0.24$, $\mu=9.115$ and $\sigma=0.556$.  
We name this skewed Gaussian model, SGE, shown as a solid black curve in Fig.~\ref{fig:fescMh}. The flat tail at the high mass end is not truncated. The value of the escape fraction at the high mass end is approximately $0.06$. We also have a variation of SGE, that we name SGET in which the flat tail is truncated at a halo mass of $10^{11.5}$ M$_\odot$. The $f_{\rm max}$  for SGET is $0.28$, slightly different from SGE to finish reionization at the same redshift, $z=6$. However, as we will see, there is only a slight variation between the models SGE and SGET as the abundance of high mass haloes is low at high redshifts.

To better understand the process of reionization by a subset of haloes and to gain insight into the mechanism of reionization by our models SGE and SGET, we create three additional models of escape fraction using the following simple Gaussian function as theoretical test cases,
\begin{equation} 
f_{\rm esc} = f_{\rm max} e^{-((x-\mu)/\sigma)^{2}}
\label{EqG}
\end{equation}
where $10^\mu$~M$_\odot$ is the mass of the specific haloes for which the escape fraction is maximum, and $\sigma$ is the width of the Gaussian distribution.
The Gaussian models can have different combinations of $\mu$, $\rm \sigma$ and  $f_{\rm max}$. However, not all such models for any random combinations of these parameters can finish reionization at redshift $6$. Therefore, we fix $f_{\rm max} = 0.5$ and run several simulations for a range of $\mu$ and $\sigma$ values on $200~\rm Mpc^3$ boxes, computed on $100^3$ grid till redshift 6. Then, we select three models calibrated to complete the reionization at $z=6$ and name them corresponding to their specific $\mu$ values. The models are named as G8, G9 and G10 corresponding to $\mu=8, 9$ and $10$. The width $\sigma$ for these models are $1.66$, $0.78$, and $0.285$, shown as red, magenta, and blue curves in Fig.~\ref{fig:fescMh} respectively.  

For all the simulations throughout this work,  we use the standard parameters for {\sc 21cmfast} from the fiducial model of \cite{Park2019}, except the escape fraction, for which we implement our models of escape fraction described above; the skewed-Gaussian models SGE and SGET (Eq.~\ref{eq_SG}) and gaussian models G8, G9, G10 (Eq.\ref{EqG}). We have also enabled inhomogeneous re-combinations in the simulations.

\subsection{Evolution of the mean neutral hydrogen fraction}
\begin{figure}
    \centering
    \includegraphics[width=\textwidth]{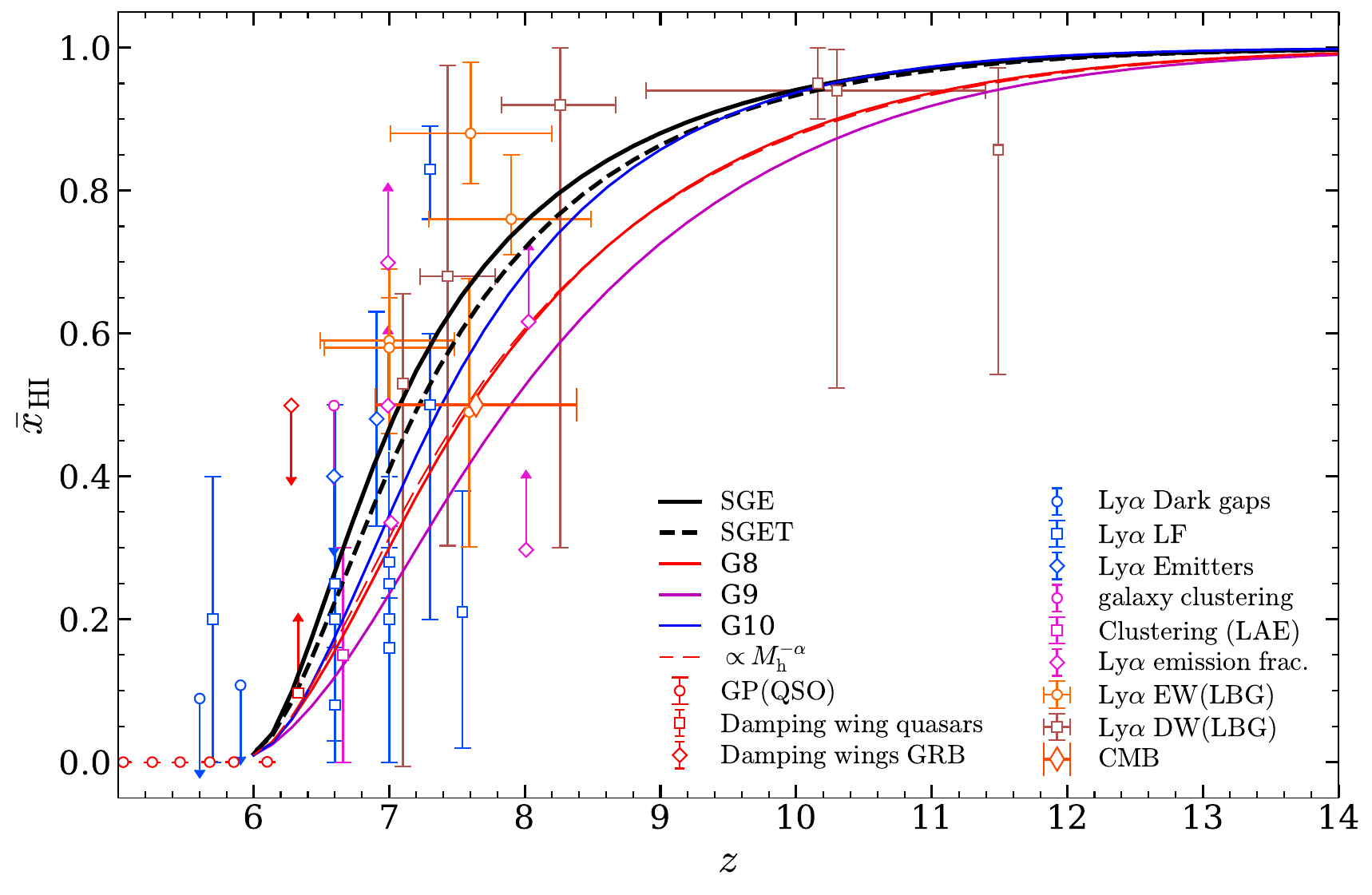}
    \caption{The evolution of the mean fraction of neutral hydrogen, $\bar{x}_{\rm HI}$, with redshift, $z$,  for our models SGE (solid black curve), SGET (dashed black curve) G8 (solid red curve), G9 (magenta curve) and G10 (blue curve). The escape fraction, $f_{\rm esc}$, as a function of halo mass, $M_{\rm h}$ for our models is shown in Fig.~\ref{fig:fescMh}. For comparison, the model with $f_{\rm esc}\propto M_{\rm h}^{-0.5}$ by \citep{Park2019} is shown as a red dashed curve. The observational data points for the measurements of neutral fraction from Lyman-$\alpha$ forest, and from quasar absorption lines and damping wings are also over-plotted \citep{Aghanim2020,Ouchi2018,Ouchi2010,Inoue2018,Morales2021,Itoh2018,Goto2021,Curtis-Lake2023,Hsiao2023,Hoag2019,Mason2019,Jung2020,Whitler2020,Bruton2023,Morishita2023,Robertson_2015,Fan2006b,McGreer2015,Schroeder2013,Mortlock2011,Totani2014,McQuinn2007a,McQuinn2007b,Ota2008,Caruana2012,Tilvi2014,Schenker2014}: red circles \citep{Fan2006b}, red square \citep{Mortlock2011}, red diamonds \citep{Totani2014}, blue circles \citep{McGreer2015}, blue square \citep{Ouchi2010, Inoue2018, Morales2021, Goto2021}, blue diamonds \citep{Ota2008, Ouchi2010}, magenta circle \citep{McQuinn2007a}, magenta square \citep{Ouchi2018}, magenta diamond \citep{Caruana2012, Schenker2014, Tilvi2014}, orange circle \citep{Hoag2019, Mason2019, Jung2020,  Whitler2020, Bruton2023, Morishita2023}, dark red square \citep{Curtis-Lake2023, Hsiao2023}, orange diamond \citep{Aghanim2020}. }
    \label{fig:lc_c0}
\end{figure}

We simulate the progress of reionization for our models of escape fraction, SGE, SGET, G8, G9, and G10, using {\sc 21cmfast} in a $500~ \rm Mpc^3$ box with a resolution of $256^3$, up to redshift 6. 
All our models and simulations are calibrated to finish the reionization at the same redshift, $z=6$. In Fig.~\ref{fig:lc_c0}, we compare the evolution of mean neutral fraction, $\bar{x}_{\rm HI}$, for our models with the observational constraints \citep{Aghanim2020,Ouchi2018,Ouchi2010,Inoue2018,Morales2021,Itoh2018,Goto2021,Curtis-Lake2023,Hsiao2023,Hoag2019,Mason2019,Jung2020,Whitler2020,Bruton2023,Morishita2023,Robertson_2015,Fan2006b,McGreer2015,Schroeder2013,Mortlock2011,Totani2014,McQuinn2007a,McQuinn2007b,Ota2008,Caruana2012,Tilvi2014,Schenker2014}. 

The reionization in our models SGE (solid black curve) and SGET (dashed black curve) progresses slowly at the beginning but then finishes rapidly at the end, compared to the models G8 and G9. The earlier studies by \citep{Naidu2020, Ishigaki2018, Finkelstein2019} found the progress of reionization to be gradual and more extended as in our models G8 and G9; however, late and steep progress as in SGE is favoured given the current observational data (see Fig.~\ref{fig:lc_c0} and  \citep{Nakane2023}).  

The reionization in the G10 model (blue curve) is similar to the SGE model, with negligible progress in the early stages but a rapid finish at the end by $z=6$, in contrast to G9 and G8. Both G8 and G9 represent low-mass haloes, and the reionization progresses earlier in them due to a higher number of ionizing photons owing to an optimum combination of halo mass and the number of haloes.

 The peak in the escape fraction for G8 occurs even at a lower halo mass than G9. However, the reionization progresses earlier in G9 compared to G8. This is due the quenching of SFR in the low mass haloes due to feedback, implemented through parameters $f_{\rm duty}$ and a constant $M_{\rm turn}$ in {\sc 21cmFAST} \citep{Park2019}.

The escape fraction in the SGE model peaks at $\approx 2\times 10^9$~M$_\odot$, but it also has a contribution from the high mass haloes of $\gtrsim 10^{10}$~M$_\odot$ (Fig.~\ref{fig:fescMh}), that plays a vital role in finishing the reionization rapidly at the end in agreement with the observational data (Fig.~\ref{fig:lc_c0}). 
Thus, the evolution of the neutral fraction for the SGE model in Fig.~\ref{fig:lc_c0} is due to a combined contribution of the lower mass ($\sim 10^9$~M$_\odot$) and the higher mass ($\gtrsim 10^{10}$~M$_\odot$) haloes. In the early stages, at $z>10$, the massive haloes are absent. Therefore, the lower mass haloes begin reionization. Subsequently, with decreasing redshift, the lower mass haloes act in parley with the upcoming high mass haloes, as they do partial reionization at smaller scales that help the high mass haloes to finish reionization swiftly at the end. In the next section, we investigate the brightness, spin, and kinetic temperature and study the spatial patchiness of reionization to understand the relative role of low and high-mass haloes in detail.

\subsection{Evolution of the kinetic, spin, and brightness temperature}
The differential brightness temperature, $\delta T_{\rm b}$, and its evolution contain significant information about the physical processes during reionization and prior to it \citep{Furlanetto2006}. The $\delta T_{\rm b}$ indicates the emission or absorption capability of neutral hydrogen. It can be written in a simple form  to understand the physics \citep{cohen17, Madau1997, Furlanetto2006, Barkana2016, Mesinger2011, Zaroubi2013}, 
\begin{equation}
    \delta T_{\rm b} = 26.8 x_{\rm HI}\Bigg(\frac{1+z}{10}\Bigg)^{1/2}(1+\delta) \Bigg[1-\frac{T_{\rm CMB}}{T_{\rm s}} \Bigg] 
\end{equation}
where $\delta$ is the matter over-density, $T_{\rm CMB}$ is the cosmic microwave background (CMB) temperature.  $T_{\rm s}$ is the spin temperature, which can be written as, 
\begin{equation}
    T_{\rm s}^{-1} = \frac{T_{\rm CMB}^{-1} + x_{\rm c}T_{\rm k}^{-1}+x_{\rm \alpha}T_{\rm C}^{-1}}{1+x_{\rm c}+x_{\rm \alpha}}  \, .
\end{equation}
Here, $T_{\rm k}$ is the kinetic (gas) temperature, and $T_{\rm c}$ is the effective temperature (colour) of Ly$\alpha$ photons.  $x_{\rm c}$ and $x_{\rm \alpha}$ are the coupling coefficients for collisions and Ly$\alpha$ scattering respectively \citep{Hirata2006, Barkana2005, Wouthuysen1952, Field1958, Mesinger2011, Zaroubi2013}.

\begin{figure}
    \centering
    \includegraphics[width=\textwidth]{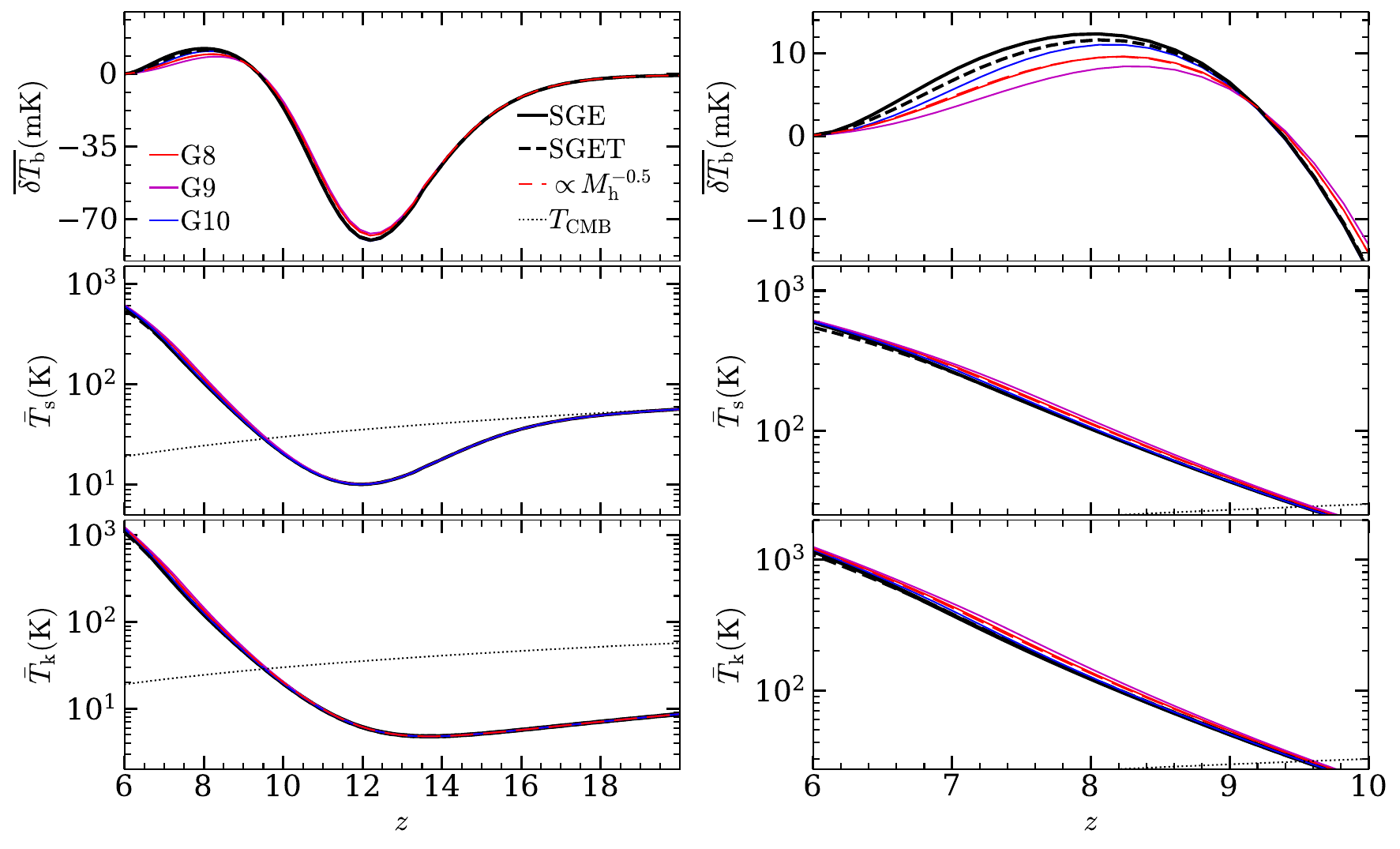}
    \caption{The evolution of the mean brightness temperature, $\overline{\delta T_{\rm b}}$ (top panels), the mean spin temperature, $\bar{T}_{\rm s}$, (middle panels) and the mean kinetic temperature, $\bar{T}_{\rm k}$ (bottom panels) for our models SGE (black solid curve), SGET (black dashed curve), G8 (red solid curve), G9 (magenta curve) and G10 (blue curve). For comparison, the model by \citep{Park2019} with $f_{\rm esc}\propto M_{\rm h}^{-0.5}$ is shown as a red dashed curve. The CMB temperature, $T_{\rm CMB}$, is shown as a dotted black line.}
    \label{fig:T_k_Ts_Tb}
\end{figure}

In Fig.~\ref{fig:T_k_Ts_Tb}, we show the evolution of the mean kinetic temperature, $\bar{T}_{\rm k}$ (bottom panel), the mean spin temperature $\bar{T}_{\rm s}$ (middle panel) and the mean brightness temperature, $\overline{\delta T}_{\rm b}$ (top panel), for our models of the escape fraction as a function of the halo mass. All the models lead to approximately the same behaviour since the $f_\star$ is the same in all these models. However, there is some difference at low redshift during reionization as the models differ in the escape fraction, $f_{\rm esc}$. The right column of plots shows specifically the behaviour during reionization (between $z=6$ and $10$).

In Fig.~\ref{fig:T_k_Ts_Tb}, the value at the peak of the brightness temperature is the highest for the model SGE (black curve), followed by G10, G8, and then G9. Intriguingly, model G9 lies below both model G8 and the fiducial model of \citep{Park2019}, similar to the trends seen in the evolution of the neural fraction. The brightness temperature depends on the neutral fraction and the kinetic temperature. The kinetic temperature shows little variation for the models as it is insensitive to the variations in the escape fraction. 

The brightness temperature for the three Gaussian models shows little difference compared to the fiducial model of \citep{Park2019}, with G9 slightly below it and G10 slightly above. The G8 model almost overlaps the model by \citep{Park2019}, which is not surprising as the escape fraction curves for these two models almost overlap (Fig.~\ref{fig:fescMh}), and both are dominated by the lowest mass haloes.

The models SGE and SGET exhibit a significant contrast from the other models, and the brightness temperature is maximum for the SGE. The peak position is almost similar for all the models, roughly at redshift $8$. However, there is a minor variation, with the peak for SGE occurring slightly later, for G9 occurring slightly earlier, and the other models peaking between the two. There is an interesting inflection point at $z\approx 9.2$ at which all the models have the same brightness temperature, as the effects of the evolution of ionized fraction and heating perfectly cancel each other at that redshift.  

The model SGE shows similarities with G10,  both in the evolution of the ionized fraction (Fig.~\ref{fig:lc_c0}) and in the evolution of the brightness temperature (Fig.~\ref{fig:T_k_Ts_Tb}), as SGE also has a significant contribution from the high mass haloes. The contribution of the massive haloes ($\gtrsim 10^{10}$~M$_\odot$) to SGE (the flat tail in Fig.~\ref{fig:fescMh}) augments the contribution from the lower mass haloes and is vital for speeding up the reionization in later stages. Thus, the trends in the evolution of $\overline{\delta T_{\rm b}}$ such as the highest values for SGE and G10, can be understood as a direct consequence of the late progress of reionization in these models that allows the Universe to remain $\approx 80\%$ neutral till $z=8$ (Fig.~\ref{fig:lc_c0}). Similarly, the lowest $\overline{\delta T_{\rm b}}$ in the case of G9 is due to the early progress of reionization in this model, as it represents low mass haloes.

\subsection{Patchiness of reionization}
\begin{figure}
    \centering
    \includegraphics[width=\textwidth]{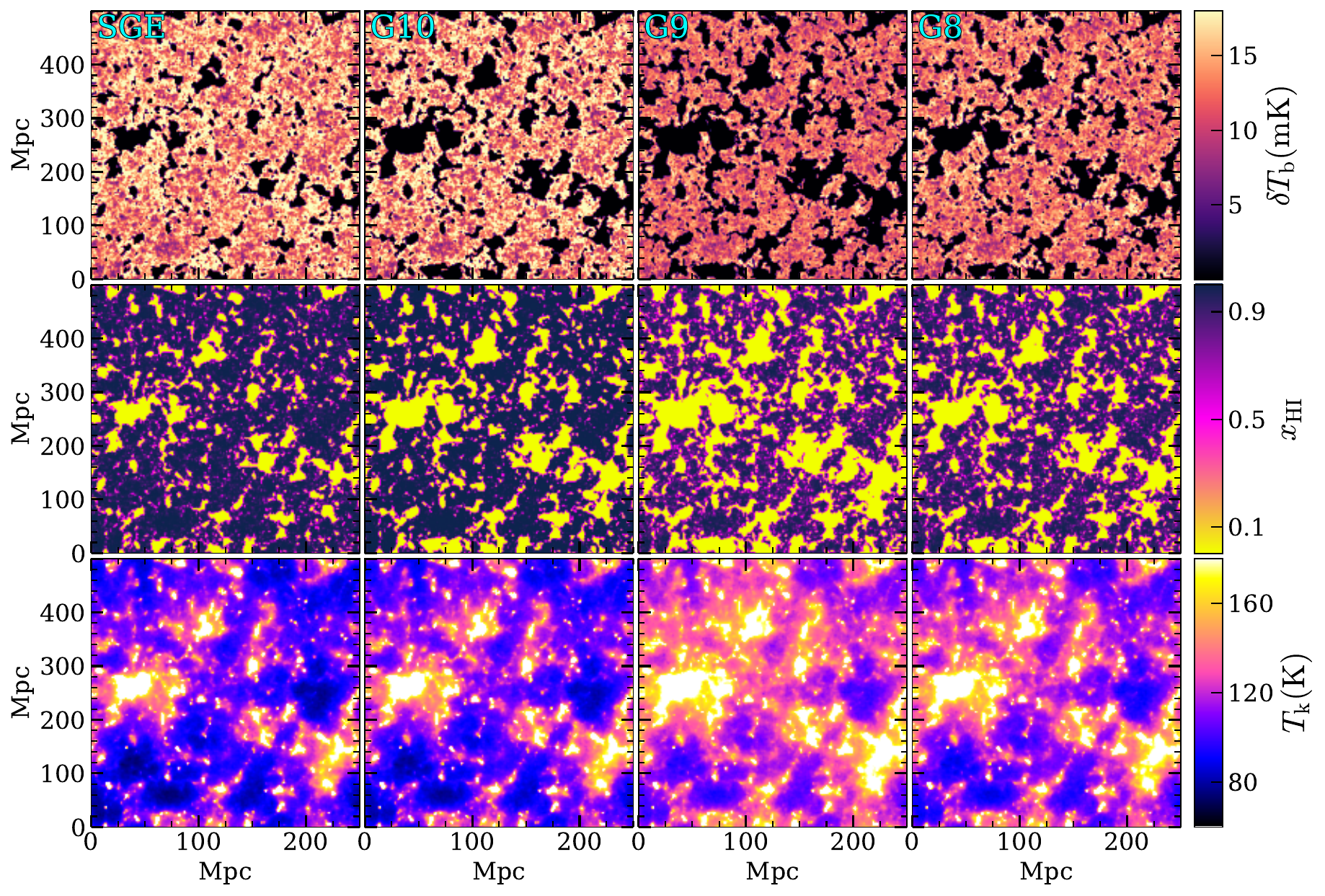}
    \caption{The 2D spatial maps showing the patchiness of reionization  at redshift $z=8$, for our models SGE, G10, G9 and G8 from left to right  respectively.}
    \label{fig:map}
\end{figure}
\begin{figure}
    \centering
    \includegraphics[width = \textwidth]{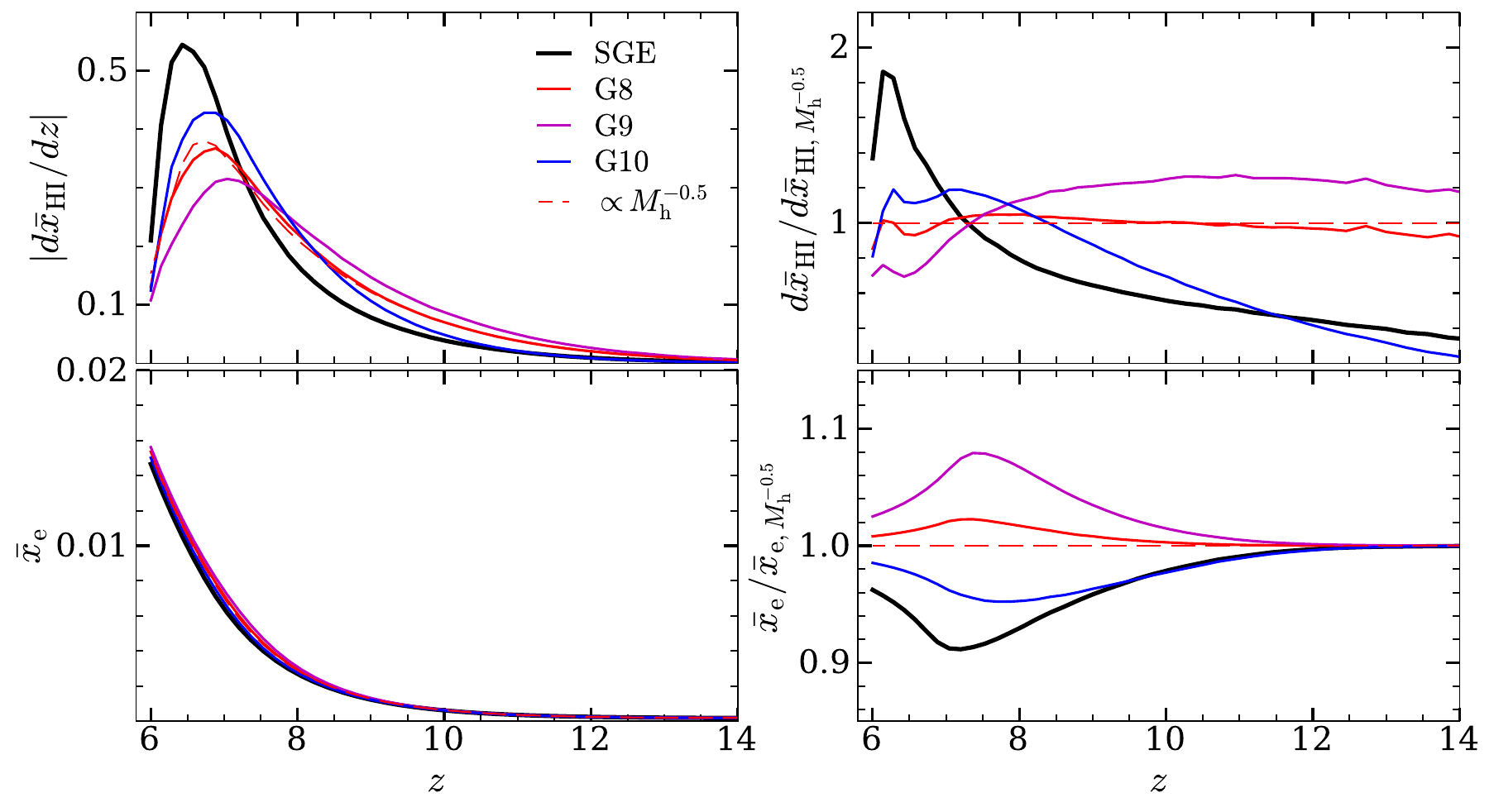}
    \caption{The evolution of the slope of $\bar{x}_{\rm HI}$ with redshift, $|d\bar{x}_{\rm HI}/dz|$ (top panels), and evolution of the mean electron fraction in the neutral regions, $\bar{x}_{\rm e}$ (bottom panels), for our models SGE (black solid curve), G8 (red solid curve), G9 (magenta curve) and G10 (blue curve). The trend of  $|d\bar{x}_{\rm HI}/dz|$ and $\bar{x}_{\rm e}$ normalized to the fiducial model of \citep{Park2019} is shown in the top right panel and in the bottom right panel respectively. }
    \label{fig:xe_dxH}
\end{figure}

In Fig.~\ref{fig:map}, we further investigate the variations in the progress of reionization by analyzing the spatial maps or the patchiness of $T_{\rm b}$,  $x_{\rm H}$, and $T_{\rm k}$, at redshift $8$. The models SGE and G9 represent two contrasting versions of reionization, with G10 and G8 bridging the gap. The kinetic temperature is almost the same in all our models, though SGE has a slightly lower value than G9, and the other models are between the two (bottom panels). However, the fact that the SGE has plenty of neutral gas (middle panels) compensates for a marginally lower kinetic temperature and results in the highest brightness temperature (top panels).

The impact of the low mass haloes in the model SGE was neither apparent in the neutral fraction's evolution nor in the brightness temperature's evolution. However, in the spatial maps (Fig.~\ref{fig:map}), the effect of low mass $\sim 10^9$~M$_\odot$ haloes to SGE becomes clear. The model SGE shows visibly more ionization at smaller scales in comparison to G10 due to the contribution from $\sim 10^9$~M$_\odot$ haloes. The large neutral regions in SGE are not entirely neutral as in G10. Instead, they are porous due to the ionization by the low-mass haloes.

In the early stages, ionization is highest in the model G9 because the Universe gets a significant number of ionizing photons early on from numerous $10^9$~M$_\odot$ haloes. However, towards the end, the massive haloes are more efficient in finishing the reionization than the low mass haloes, as is seen in the models G10 and SGE.

At $z=8$, SGE is the least ionized model, which is also reflected in the patchiness maps. The SGE has signs of reionization at smaller scales due to contribution from low-mass haloes and also has a few large ionized regions due to a few but powerful massive sources. Nevertheless, in SGE the Universe is mostly neutral at redshift 8. However, below redshift 8, it undergoes a rapid transition as an increasingly higher number of massive galaxies develop rapidly from redshift 8 to 6, and, together with the lower mass haloes, they flood the Universe with ionizing photons to finish reionization at redshift 6. 

In Fig.~\ref{fig:xe_dxH}, we show the slope of $\bar{x}_{\rm HI}$ as a function of redshift. There is a sharp upturn in the slope of $d\bar{x}_{\rm HI}/dz$  (the top panels in Fig.~\ref{fig:xe_dxH}). In fact, the rise in the black curve corresponding to SGE is staggering, revealing a drastic change in the SGE model between redshift $8$ and $6.5$, akin to a phase transition. All of a sudden, there is rapid progress in reionization within a short period, and that, in fact, stands out in comparison to other models as SGE has a special push to the reionization by the high-mass galaxies. 

In the lower panels of Fig.~\ref{fig:xe_dxH}, we also show the trends of mean electron fraction within the neutral regions, $\bar{x}_e$, which is correlated with the kinetic temperature \citep{Mesinger2011, Mesinger2013, Park2019}.  The trends of $\bar{x}_e$ reflect the behaviour of $\bar{T}_{\rm k}$, as the black curve for SGE is the lowest. In contrast, the magenta curve for G9 is the highest. The magnitude of the difference is only $\sim 10$\%, therefore it has a negligible effect on the brightness temperature, that too is negated and overpowered by the trends in the neutral fraction (upper panels).

The model G10 also exhibits an increase in the slope of $\bar{x}_{\rm HI}$ with decreasing redshift; however, it is not as staggering as in SGE because the G10 model only features ($\sim 10^{10}$~M$_\odot$) haloes. A combination of the lower ($\sim 10^{9}$~M$_\odot$) and higher mass ($>10^{10}$~M$_\odot$) haloes is at work in SGE. The haloes of $\sim 10^{9}$~M$_\odot$ are not affected by the supernovae and photoheating feedback implemented through a constant $M_{\rm turn}$  \citep{Park2019}  and they cause early ionization uniformly at small scales, which has a catalytic effect that amplifies the contribution of the high mass ($>10^{10}$~M$_\odot$) haloes in causing steep progress of reionization in later stages in the model SGE.

The model SGE is, therefore, peculiar in the evolution of the neutral fraction. The $\bar{x}_{\rm HI}$ in it remains high till late but then diminishes steeply at a very rapid rate at $z < 8$ in comparison to other models (top panels of Fig.~\ref{fig:xe_dxH}),  which is desirable according to recent studies \citep{Nakane2023} as it leads to a good agreement with the observational data on $\bar{x}_{\rm HI}$ (Fig.~\ref{fig:lc_c0}) and also leads to a prominent peak in the brightness temperature (Fig.~\ref{fig:T_k_Ts_Tb}).

\subsection{21 cm power spectrum}
\begin{figure}
    \centering
    \includegraphics[width=\textwidth]{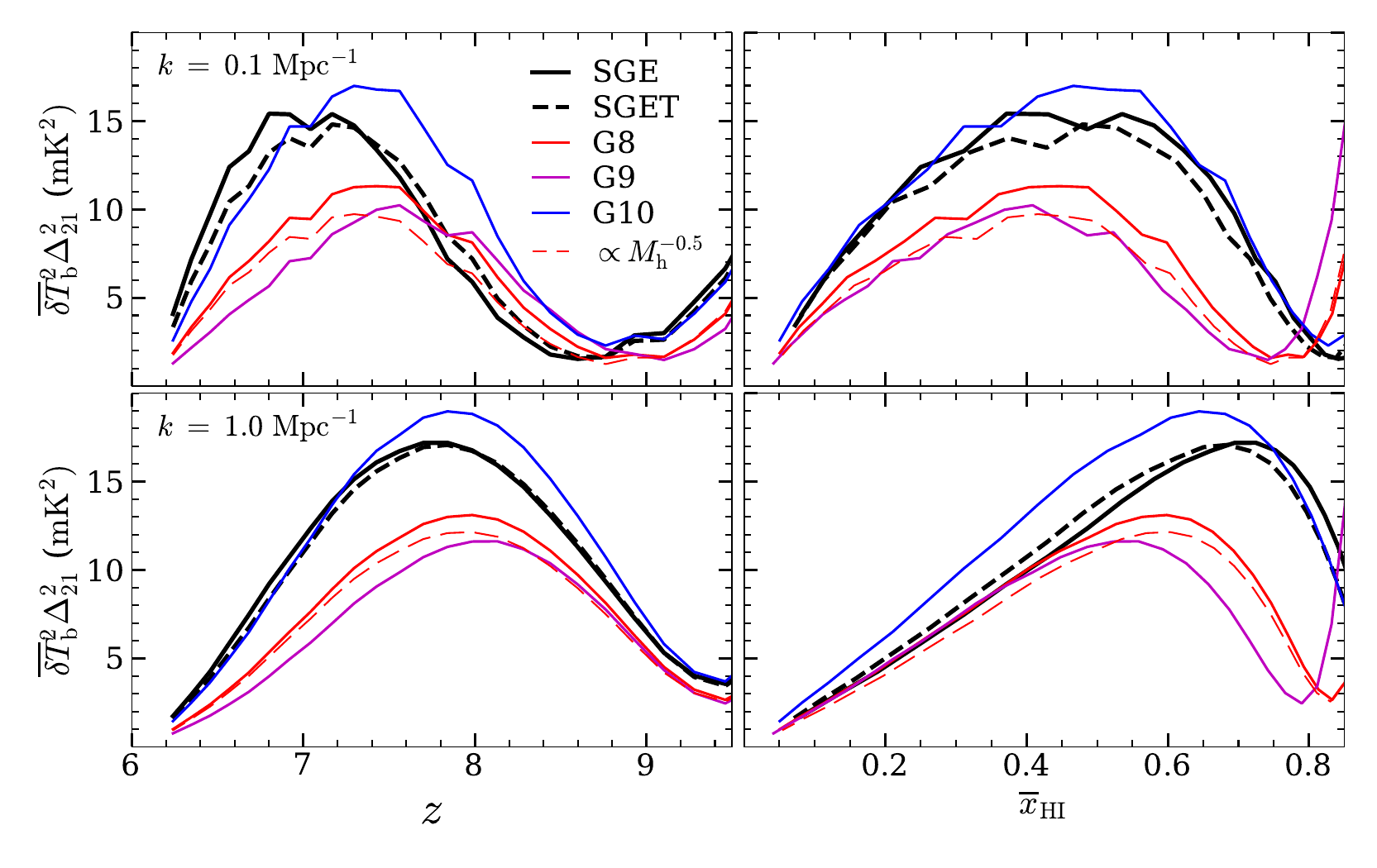}
    \caption{The evolution of the power spectrum at the 21 cm wavelength, $\overline{\delta T^2_{\rm b}}\Delta_{21}^2$, with redshift, $z$ (left panels), and with the mean neutral fraction, $\bar{x}_{\rm HI}$ (right panels); at $k=0.1$~Mpc$^{-1}$ (top panels), and at $k=1$~Mpc$^{-1}$ (bottom panels); for our models of escape fraction: SGE (black solid curve), SGET (black dashed curve), G8 (red solid curve), G9 (magenta curve) and G10 (blue curve). The fiducial model by \citep{Park2019} is shown as a red dashed curve for comparison.     }
    \label{fig_power}
\end{figure}
\begin{figure}
    \centering
    \includegraphics[width=\textwidth]{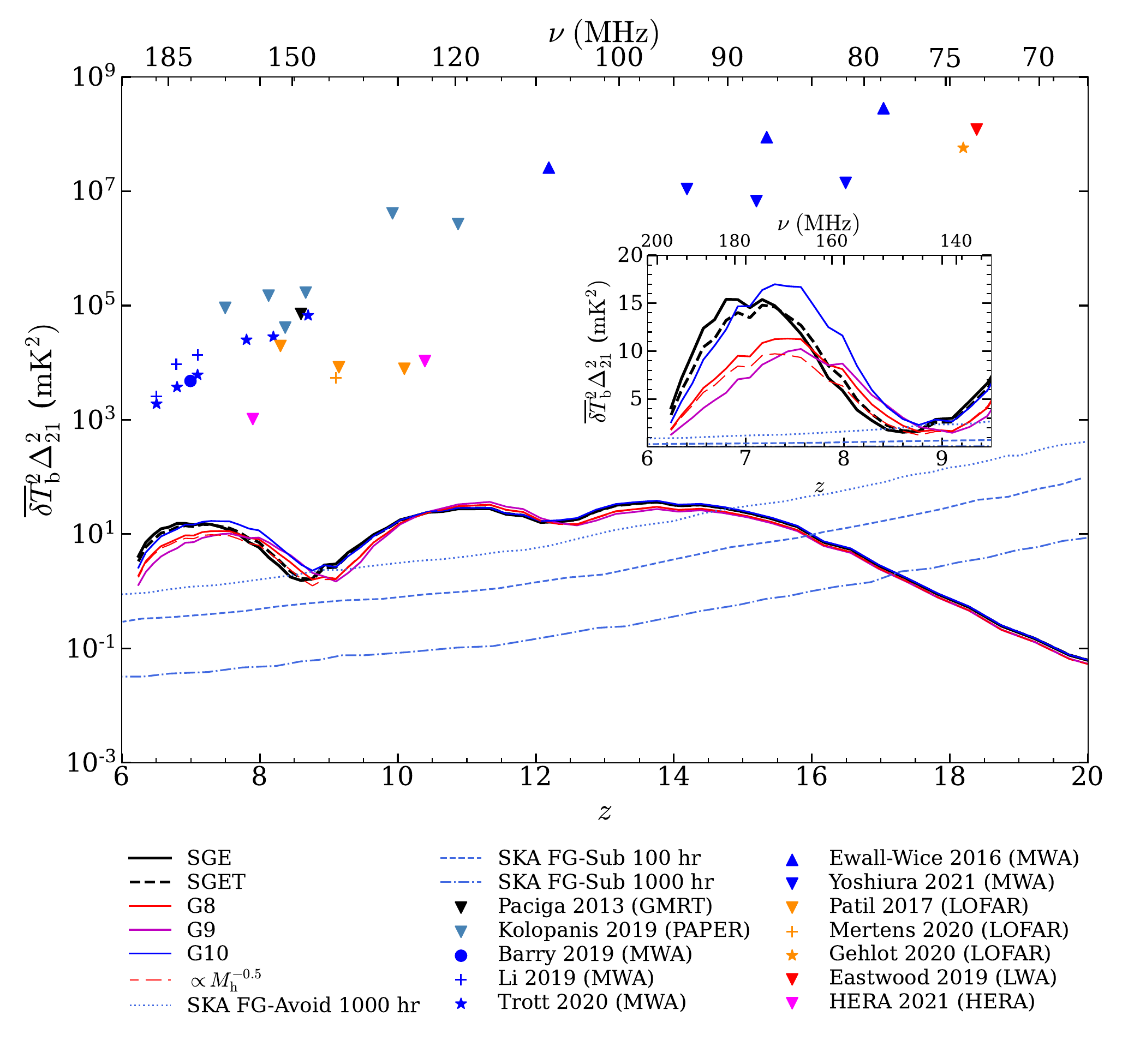}
    \caption{The evolution of the power spectrum at the 21 cm wavelength, $\overline{\delta T^2_{\rm b}}\Delta_{21}^2$, with redshift, for our models of escape fraction, $f_{\rm esc}$ as a function of halo mass, $M_{\rm h}$: SGE (black solid curve), SGET (black dashed curve), G8 (red solid curve), G9 (magenta curve) and G10 (blue curve). The fiducial model by \citep{Park2019} with $f_{\rm esc}$ is shown as a red dashed line. The upcoming Square Kilometre Array (SKA) sensitivity limits with varying integration times are shown as blue dotted, dashed, and dash-dotted lines \cite{Barry2022}. The current sensitivity limits of the existing surveys are shown as symbols \cite{Abdurashidova2022a, Abdurashidova2022b, Ewall-Wice2016, Barry2019, Li2019, Trott2020, Yoshiura2021, Kolopanis2019, Patil2017, Mertens2020, Gehlot2020, Paciga2013, HERA2022, Eastwood2019}.}
    \label{fig:radio_obs}
\end{figure}

The fluctuations in the brightness temperature impact the power spectrum at 21~cm. In Fig.~\ref{fig_power}, we show the power spectrum of neutral hydrogen at small scales (bottom panels) and large scales (top panels). As expected, the power is the highest for our models SGE and G10 since they both maintain a neutral Universe till late. The models G9 and G8 have almost the same power as in the fiducial model of \citep{Park2019}.

The model SGE has slightly lower power than the model G10, and the difference in the power spectrum is higher at the smaller scales (bottom panels in Fig.~\ref{fig_power}). This is again a consequence of the fact that the neutral regions in SGE are porous such that they have some degree of ionization at small scales due to the low-mass haloes. The peak in power at large scales in the model SGE occurs at a lower redshift than other models (top left panel). It is due to the variation in the progress of reionization with redshift across models. Indeed, when the power spectrum is plotted as a function of the $\bar{x}_{\rm HI}$,  the peak occurs almost at the same $\bar{x}_{\rm HI}$ in all our models (top right panel).

In Fig.~\ref{fig:radio_obs}, we compare the power spectrum of the expected 21 cm signal for our model SGE, G10, G9, and G8, with the sensitivities of ongoing and upcoming observational surveys \citep{Barry2022, Abdurashidova2022a, Abdurashidova2022b, Ewall-Wice2016, Barry2019, Li2019, Trott2020, Yoshiura2021, Kolopanis2019, Patil2017, Mertens2020, Gehlot2020, Paciga2013, HERA2022, Eastwood2019}. As expected, the power at low redshift $z\le 8$ is the highest for our models SGE and G10. The peak in the power spectrum in the models G8 and G9 occurs roughly at $z\approx 7.5$, similar to that in \citep{Park2019}. On the other hand, for our models, SGE and G10, the maxima occur at $z\approx 6.8$ corresponding to $\nu \approx 180$~MHz for the redshifted 21 cm line. The maxima is prominent and has a value $\overline{\delta T^2_{\rm b}}\Delta_{21}^2 \approx 15$~\rm mK$^2$.

The maxima in the power spectrum for our model SGE occurs at a lower redshift than in the fiducial model by \citep{Park2019}.  The model by \citep{Park2019} represents low mass haloes and has a different reionization morphology than the model SGE, but it is almost similar to our model G8. However, the SGE is physically motivated by the nature of star formation in first galaxies, and it successfully reproduces a late and rapid reionization desired by the current observational data \citep{Nakane2023}. We conclude that the Universe was reionized by $\gtrsim 10^9$~M$_\odot$ haloes as shown by the escape fraction curve in Fig.~\ref{fig:fescMh} for our model pick SGE.

\section{Summary}
The shape of the halo mass function \citep{Schechter1976, Sheth1999, Jenkins2001} implies that the number of haloes increases with decreasing mass. Galaxy luminosity functions also rise towards the faint end \citep{Bouwens2015, Bouwens2023}. Integration of the luminosity function weighted by the luminosity, and considering each galaxy contributes a fixed fraction of its ionizing photons, leads to the viewpoint that the main contribution comes from the faint low-mass galaxies due to their numerical superiority \citep{Robertson_2015, Mesinger2011, Finkelstein2019, Atek2024}. This viewpoint has been challenged by the models based on the recent detection of ionizing photons from the starburst galaxies \citep{Izotov2016, Schaerer2019}, that have abnormally high SFR,  or more appropriately, the high values of $\Sigma_{\rm SFR}$ and sSFR \citep{Sharma2017, Calabro2024}. Such galaxies are prone to violent bursts and outflows, which facilitate the escape of photons. A natural conclusion from this line of study is that the Universe was reionized by the bursty brighter galaxies, not by the faintest ones \citep{Sharma2016}. The term `brighter' is prone to an interpretation to mean `only the most massive' on the assumption of a correlation without scatter between the SFR and mass, even at the highest redshifts $z\ge 6$. 

Though the high-mass haloes host the brightest galaxies, the picture is more complex at the low-mass end. A significant scatter in the SFR and $M_{\rm UV}$ is observed at the low mass end. The pre-reionization low mass haloes host the expected faint galaxies, but they also host the `brighter' bursty galaxies (Fig.~\ref{fig_first}). Such bursty galaxies within low-mass haloes have $M_{\rm UV}$ between  $-17$ and $-21$ \citep{Robertson2023a, Looser2023, Bunker2023, Jones2023, Curti2023}, which is comparable to the brightness of most massive haloes ($M_{\rm UV}\lesssim -21$), and is significantly higher than the expected faint magnitudes ($M_{\rm UV}\gtrsim -17$) for such low mass haloes \citep{Atek2024}. We find that an appropriate parameter to identify the bursty galaxies that reionized the Universe is the specific star formation rate, sSFR. The bursty low-mass galaxies from JADES \citep{Robertson2023a, Looser2023}, and a few slightly higher mass galaxies from CEERS \citep{Nakane2023, Nakajima2023, Jung2023, Haro2023, Harikane2024, Zitrin2015, Fujimoto2023, Tang2023, Larson2023, Sanders2023} and GLASS \citep{Nakajima2023} have above-normal sSFR~$> 10^{-2}$~Myr$^{-1}$.

We further use the cosmological simulation {\sc eagle} \citep{Schaye2015} to investigate the haloes that host bursty galaxies with above-normal sSFR at $z\ge 6$ to identify the haloes that reionized the Universe. The scatter plot of $\rm sSFR$ as a function of the halo mass for the galaxies in the {\sc eagle} simulation (Fig.~\ref{fig_SFR_Mh}) compared with the JADES data \citep{Looser2023} show that the galaxies exhibit a peak in the sSFR at a halo mass of $\sim 10^9$~M$_\odot$.

By considering haloes corresponding to the peak with above-normal sSFR to have the highest escape fraction, we develop models of the escape fraction as a function of the halo mass. A skewed Gaussian function with a flat tail (Eq.\ref{eq_SG}) is the preferred model that accurately represents the burstiness of galaxies at $z\ge 6$, which we call SGE (and SGET with a slight variation). We have three additional models, G8, G9, and G10, represented by the standard Gaussian function with variation in the location of the peak and width (Eq.\ref{EqG}). The maximum escape fraction for our main model SGE is $0.25$ for the $\sim 10^9$~M$_\odot$ haloes and $0.06$ for the massive haloes.

We implement our models of escape fraction in the publicly available code {\sc 21cmfast} \citep{Mesinger2011, Park2019} to study the progress of reionization and the statistical signal at 21 cm wavelength. The evolution of the mean ionized fraction for our main model SGE agrees well with the observational measurements \cite{Aghanim2020, Ouchi2018, Ouchi2010, Inoue2018, Morales2021, Itoh2018, Goto2021, Curtis-Lake2023, Hsiao2023, Hoag2019, Mason2019, Jung2020, Whitler2020, Bruton2023, Morishita2023, Robertson_2015, Fan2006b, McGreer2015, Schroeder2013, Mortlock2011, Totani2014, McQuinn2007a, McQuinn2007b, Ota2008, Caruana2012, Tilvi2014, Schenker2014}. 
The SGE has a peak in the escape fraction at $\sim 10^9$~M$_\odot$ similar to G9, but surprisingly in the $\bar{x}_{\rm HI}$ evolution, the Gaussian model G10 lies nearest to SGE (Fig.~\ref{fig:lc_c0}). Further, in the evolution of the brightness temperature, the SGE  has the highest brightness temperature during reionization as it keeps the Universe neutral till late (Fig.~\ref{fig:T_k_Ts_Tb}). 

We further investigate this in the spatial maps of the brightness temperature and neutral fraction for our models (Fig.~\ref{fig:map}). The SGE model displays an efficient combination of two factors: first, the late but high contribution of photons from massive galaxies, and second, the factor that has a catalytic effect is the porous structure of neutral gas due to ionization by low-mass haloes. Therefore in SGE, the Universe is neutral till late (50\% neutral upto $z\approx 7$) but then reionizes rapidly, in fact almost instantaneously, akin to a phase transition (Fig.~\ref{fig:xe_dxH}). 

The $f_{\rm esc}$ for the model SGE has a flat tail towards the high mass end (Fig.~\ref{fig:fescMh}), though there seems to be a downturn in the sSFR towards the high mass end (Fig.~\ref{fig_SFR_Mh}). However, even by excluding the contribution of the haloes above $\approx 3\times 10^{11}$~M$_\odot$ as we do in our model SGET, the results do not vary much. Therefore, the SGE and SGET are the preferred models that show late rapid progress of reionization in agreement with observations (Fig.~\ref{fig:lc_c0}). 

 We conclude that the Universe was reionized by the haloes represented by the SGE model, as they host bursty galaxies with sSFR~$>10^{-2}$~Myr$^{-1}$ and UV magnitude, $M_{\rm UV}\leq -17$. As per the SGE model, the bursty galaxies reside predominantly in $\gtrsim 10^9$~M$_\odot$ haloes. A minority of them also reside in higher mass haloes, which play a vital role, particularly at later redshifts  ($z < 8$), by increasing the speed of reionization. 
In the power spectrum of 21 cm (Fig.~\ref{fig:radio_obs}), the SGE has subtle noticeable differences from the earlier models in the literature \citep{Park2019}, and also from our models G8, G9, and G10. It has a higher power at redshifts $z=6$ to $z=7.5$, with a peak at $z\approx 6.8$ corresponding to frequency $180$~MHz for the redshifted 21 cm line. The upcoming SKA will be able to test our predictions \citep{Koopmans2015, Mellema2015, Barry2022}.

\section*{Acknowledgements}
We thank an anonymous referee for insightful comments that improved the paper. NJ and MS thank the Department of Science and Technology (DST), Science and Engineering Research Board (SERB), India, for the support through research grants SRG/2022/001137 and MTR/2022/000548. The authors thank the Virgo Consortium for making their simulation data publicly available. The {\sc eagle} simulation was performed using the DiRAC-2 facility at Durham, managed by the ICC, and the PRACE Curie facility based in France at TGCC, CEA, Bruy\`eres-le-Ch\^atel. MS thanks C. Trott and the EoR group at CIRA for fruitful discussions during an ASTRO-3D fellowship at ICRAR Curtin. The authors also thank the developers of the code {\sc 21cmfast} for making their code publicly available.


\bibliographystyle{JCAP}
\bibliography{ref_21}

\providecommand{\href}[2]{#2}\begingroup\raggedright\begin{thebibliography}{100}

\bibitem{Naoz2006}
S.~{Naoz}, S.~{Noter} and R.~{Barkana}, \emph{{The first stars in the
  Universe}},
  \href{https://doi.org/10.1111/j.1745-3933.2006.00251.x}{\emph{\mnras}
  {\bfseries 373} (2006) L98}
  [\href{https://arxiv.org/abs/astro-ph/0604050}{{\ttfamily
  astro-ph/0604050}}].

\bibitem{Loeb2001}
A.~{Loeb} and R.~{Barkana}, \emph{{The Reionization of the Universe by the
  First Stars and Quasars}},
  \href{https://doi.org/10.1146/annurev.astro.39.1.19}{\emph{\araa} {\bfseries
  39} (2001) 19} [\href{https://arxiv.org/abs/astro-ph/0010467}{{\ttfamily
  astro-ph/0010467}}].

\bibitem{Rauch1998}
M.~{Rauch}, \emph{{The Lyman Alpha Forest in the Spectra of QSOs}},
  \href{https://doi.org/10.1146/annurev.astro.36.1.267}{\emph{\araa} {\bfseries
  36} (1998) 267} [\href{https://arxiv.org/abs/astro-ph/9806286}{{\ttfamily
  astro-ph/9806286}}].

\bibitem{Fan2003}
X.~{Fan}, M.A.~{Strauss}, D.P.~{Schneider}, R.H.~{Becker}, R.L.~{White},
  Z.~{Haiman} et~al., \emph{{A Survey of z>5.7 Quasars in the Sloan Digital Sky
  Survey. II. Discovery of Three Additional Quasars at z>6}},
  \href{https://doi.org/10.1086/368246}{\emph{\aj} {\bfseries 125} (2003) 1649}
  [\href{https://arxiv.org/abs/astro-ph/0301135}{{\ttfamily
  astro-ph/0301135}}].

\bibitem{Fan2006}
X.~{Fan}, M.A.~{Strauss}, G.T.~{Richards}, J.F.~{Hennawi}, R.H.~{Becker},
  R.L.~{White} et~al., \emph{{A Survey of z>5.7 Quasars in the Sloan Digital
  Sky Survey. IV. Discovery of Seven Additional Quasars}},
  \href{https://doi.org/10.1086/500296}{\emph{\aj} {\bfseries 131} (2006) 1203}
  [\href{https://arxiv.org/abs/astro-ph/0512080}{{\ttfamily
  astro-ph/0512080}}].

\bibitem{Aghanim2020}
{Planck Collaboration}, N.~{Aghanim}, Y.~{Akrami}, M.~{Ashdown}, J.~{Aumont},
  C.~{Baccigalupi} et~al., \emph{{Planck 2018 results. VI. Cosmological
  parameters}}, \href{https://doi.org/10.1051/0004-6361/201833910}{\emph{\aap}
  {\bfseries 641} (2020) A6}
  [\href{https://arxiv.org/abs/1807.06209}{{\ttfamily 1807.06209}}].

\bibitem{Barkana2016}
R.~{Barkana}, \emph{{The rise of the first stars: Supersonic streaming,
  radiative feedback, and 21-cm cosmology}},
  \href{https://doi.org/10.1016/j.physrep.2016.06.006}{\emph{\physrep}
  {\bfseries 645} (2016) 1} [\href{https://arxiv.org/abs/1605.04357}{{\ttfamily
  1605.04357}}].

\bibitem{Sharma2016}
M.~{Sharma}, T.~{Theuns}, C.~{Frenk}, R.~{Bower}, R.~{Crain}, M.~{Schaller}
  et~al., \emph{{The brighter galaxies reionized the Universe}},
  \href{https://doi.org/10.1093/mnrasl/slw021}{\emph{\mnras} {\bfseries 458}
  (2016) L94} [\href{https://arxiv.org/abs/1512.04537}{{\ttfamily
  1512.04537}}].

\bibitem{Sharma2017}
M.~{Sharma}, T.~{Theuns}, C.~{Frenk}, R.G.~{Bower}, R.A.~{Crain}, M.~{Schaller}
  et~al., \emph{{Winds of change: reionization by starburst galaxies}},
  \href{https://doi.org/10.1093/mnras/stx578}{\emph{\mnras} {\bfseries 468}
  (2017) 2176} [\href{https://arxiv.org/abs/1606.08688}{{\ttfamily
  1606.08688}}].

\bibitem{Sokasian2004}
A.~{Sokasian}, N.~{Yoshida}, T.~{Abel}, L.~{Hernquist} and V.~{Springel},
  \emph{{Cosmic reionization by stellar sources: population III stars}},
  \href{https://doi.org/10.1111/j.1365-2966.2004.07636.x}{\emph{\mnras}
  {\bfseries 350} (2004) 47}
  [\href{https://arxiv.org/abs/astro-ph/0307451}{{\ttfamily
  astro-ph/0307451}}].

\bibitem{Haardt2012}
F.~{Haardt} and P.~{Madau}, \emph{{Radiative Transfer in a Clumpy Universe. IV.
  New Synthesis Models of the Cosmic UV/X-Ray Background}},
  \href{https://doi.org/10.1088/0004-637X/746/2/125}{\emph{\apj} {\bfseries
  746} (2012) 125} [\href{https://arxiv.org/abs/1105.2039}{{\ttfamily
  1105.2039}}].

\bibitem{Robertson2013}
B.E.~{Robertson}, S.R.~{Furlanetto}, E.~{Schneider}, S.~{Charlot},
  R.S.~{Ellis}, D.P.~{Stark} et~al., \emph{{New Constraints on Cosmic
  Reionization from the 2012 Hubble Ultra Deep Field Campaign}},
  \href{https://doi.org/10.1088/0004-637X/768/1/71}{\emph{\apj} {\bfseries 768}
  (2013) 71} [\href{https://arxiv.org/abs/1301.1228}{{\ttfamily 1301.1228}}].

\bibitem{Madau2015}
P.~{Madau} and F.~{Haardt}, \emph{{Cosmic Reionization after Planck: Could
  Quasars Do It All?}},
  \href{https://doi.org/10.1088/2041-8205/813/1/L8}{\emph{\apjl} {\bfseries
  813} (2015) L8} [\href{https://arxiv.org/abs/1507.07678}{{\ttfamily
  1507.07678}}].

\bibitem{Mitra2018}
S.~{Mitra}, T.R.~{Choudhury} and A.~{Ferrara}, \emph{{Cosmic reionization after
  Planck II: contribution from quasars}},
  \href{https://doi.org/10.1093/mnras/stx2443}{\emph{\mnras} {\bfseries 473}
  (2018) 1416} [\href{https://arxiv.org/abs/1606.02719}{{\ttfamily
  1606.02719}}].

\bibitem{Robertson_2015}
B.E.~Robertson, R.S.~Ellis, S.R.~Furlanetto and J.S.~Dunlop, \emph{Cosmic
  reionization and early star-forming galaxies: A joint analysis of new
  constraints from planck and the hubble space telescope},
  \href{https://doi.org/10.1088/2041-8205/802/2/L19}{\emph{The Astrophysical
  Journal Letters} {\bfseries 802} (2015) L19}.

\bibitem{Bouwens_2015}
R.J.~Bouwens, G.D.~Illingworth, P.A.~Oesch, J.~Caruana, B.~Holwerda, R.~Smit
  et~al., \emph{Reionization after planck: The derived growth of the cosmic
  ionizing emissivity now matches the growth of the galaxy uv luminosity
  density*}, \href{https://doi.org/10.1088/0004-637X/811/2/140}{\emph{The
  Astrophysical Journal} {\bfseries 811} (2015) 140}.

\bibitem{Bromm2009}
V.~{Bromm}, N.~{Yoshida}, L.~{Hernquist} and C.F.~{McKee}, \emph{{The formation
  of the first stars and galaxies}},
  \href{https://doi.org/10.1038/nature07990}{\emph{\nat} {\bfseries 459} (2009)
  49} [\href{https://arxiv.org/abs/0905.0929}{{\ttfamily 0905.0929}}].

\bibitem{Barkana2001}
R.~{Barkana} and A.~{Loeb}, \emph{{In the beginning: the first sources of light
  and the reionization of the universe}},
  \href{https://doi.org/10.1016/S0370-1573(01)00019-9}{\emph{\physrep}
  {\bfseries 349} (2001) 125}
  [\href{https://arxiv.org/abs/astro-ph/0010468}{{\ttfamily
  astro-ph/0010468}}].

\bibitem{Shibuya2015}
T.~{Shibuya}, M.~{Ouchi} and Y.~{Harikane}, \emph{{Morphologies of
  {\ensuremath{\sim}}190,000 Galaxies at z = 0-10 Revealed with HST Legacy
  Data. I. Size Evolution}},
  \href{https://doi.org/10.1088/0067-0049/219/2/15}{\emph{\apjs} {\bfseries
  219} (2015) 15} [\href{https://arxiv.org/abs/1503.07481}{{\ttfamily
  1503.07481}}].

\bibitem{Furlong2015}
M.~{Furlong}, R.G.~{Bower}, T.~{Theuns}, J.~{Schaye}, R.A.~{Crain},
  M.~{Schaller} et~al., \emph{{Evolution of galaxy stellar masses and star
  formation rates in the EAGLE simulations}},
  \href{https://doi.org/10.1093/mnras/stv852}{\emph{\mnras} {\bfseries 450}
  (2015) 4486} [\href{https://arxiv.org/abs/1410.3485}{{\ttfamily 1410.3485}}].

\bibitem{Finkelstein2015}
S.L.~{Finkelstein}, J.~{Ryan}, Russell~E., C.~{Papovich}, M.~{Dickinson},
  M.~{Song}, R.S.~{Somerville} et~al., \emph{{The Evolution of the Galaxy
  Rest-frame Ultraviolet Luminosity Function over the First Two Billion
  Years}}, \href{https://doi.org/10.1088/0004-637X/810/1/71}{\emph{\apj}
  {\bfseries 810} (2015) 71} [\href{https://arxiv.org/abs/1410.5439}{{\ttfamily
  1410.5439}}].

\bibitem{cohen17}
A.~{Cohen}, A.~{Fialkov}, R.~{Barkana} and M.~{Lotem}, \emph{{Charting the
  parameter space of the global 21-cm signal}},
  \href{https://doi.org/10.1093/mnras/stx2065}{\emph{\mnras} {\bfseries 472}
  (2017) 1915} [\href{https://arxiv.org/abs/1609.02312}{{\ttfamily
  1609.02312}}].

\bibitem{endsley2023}
R.~Endsley, D.P.~Stark, L.~Whitler, M.W.~Topping, B.D.~Johnson, B.~Robertson
  et~al., \emph{The star-forming and ionizing properties of dwarf z\~{} 6-9
  galaxies in jades: Insights on bursty star formation and ionized bubble
  growth}, {\emph{arXiv preprint arXiv:2306.05295} (2023) }.

\bibitem{Nakane2023}
M.~{Nakane}, M.~{Ouchi}, K.~{Nakajima}, Y.~{Harikane}, Y.~{Ono}, H.~{Umeda}
  et~al., \emph{{Ly$\alpha$ Emission at $z=7-13$: Clear Ly$\alpha$ Equivalent
  Width Evolution Indicating the Late Cosmic Reionization History}},
  \href{https://doi.org/10.48550/arXiv.2312.06804}{\emph{arXiv e-prints} (2023)
  arXiv:2312.06804} [\href{https://arxiv.org/abs/2312.06804}{{\ttfamily
  2312.06804}}].

\bibitem{Simmonds2024}
C.~{Simmonds}, S.~{Tacchella}, K.~{Hainline}, B.D.~{Johnson}, W.~{McClymont},
  B.~{Robertson} et~al., \emph{{Low-mass bursty galaxies in JADES efficiently
  produce ionizing photons and could represent the main drivers of
  reionization}}, \href{https://doi.org/10.1093/mnras/stad3605}{\emph{\mnras}
  {\bfseries 527} (2024) 6139}
  [\href{https://arxiv.org/abs/2310.01112}{{\ttfamily 2310.01112}}].

\bibitem{Atek2015}
H.~{Atek}, J.~{Richard}, M.~{Jauzac}, J.-P.~{Kneib}, P.~{Natarajan},
  M.~{Limousin} et~al., \emph{{Are Ultra-faint Galaxies at z = 6-8 Responsible
  for Cosmic Reionization? Combined Constraints from the Hubble Frontier Fields
  Clusters and Parallels}},
  \href{https://doi.org/10.1088/0004-637X/814/1/69}{\emph{\apj} {\bfseries 814}
  (2015) 69} [\href{https://arxiv.org/abs/1509.06764}{{\ttfamily 1509.06764}}].

\bibitem{Bouwens2017}
R.J.~{Bouwens}, P.A.~{Oesch}, G.D.~{Illingworth}, R.S.~{Ellis} and
  M.~{Stefanon}, \emph{{The z {\ensuremath{\sim}} 6 Luminosity Function Fainter
  than -15 mag from the Hubble Frontier Fields: The Impact of Magnification
  Uncertainties}}, \href{https://doi.org/10.3847/1538-4357/aa70a4}{\emph{\apj}
  {\bfseries 843} (2017) 129}
  [\href{https://arxiv.org/abs/1610.00283}{{\ttfamily 1610.00283}}].

\bibitem{Atek2024}
H.~{Atek}, I.~{Labbé}, L.J.~{Furtak}, I.~{Chemerynska}, S.~{Fujimoto},
  D.J.~{Setton} et~al., \emph{{Most of the photons that reionized the Universe
  came from dwarf galaxies}},
  \href{https://doi.org/10.1038/s41586-024-07043-6}{\emph{\nat} {\bfseries 626}
  (2024) 975}.

\bibitem{Sharma2018}
M.~{Sharma}, T.~{Theuns} and C.~{Frenk}, \emph{{The duration of reionization
  constrains the ionizing sources}},
  \href{https://doi.org/10.1093/mnrasl/sly052}{\emph{\mnras} {\bfseries 477}
  (2018) L111} [\href{https://arxiv.org/abs/1712.06619}{{\ttfamily
  1712.06619}}].

\bibitem{Qin2023}
Y.~{Qin}, S.~{Balu} and J.S.B.~{Wyithe}, \emph{{Implications of z
  {\ensuremath{\gtrsim}} 12 JWST galaxies for galaxy formation at high
  redshift}}, \href{https://doi.org/10.1093/mnras/stad2448}{\emph{\mnras}
  {\bfseries 526} (2023) 1324}
  [\href{https://arxiv.org/abs/2305.17959}{{\ttfamily 2305.17959}}].

\bibitem{Looser2023}
T.J.~{Looser}, F.~{D'Eugenio}, R.~{Maiolino}, S.~{Tacchella}, M.~{Curti},
  S.~{Arribas} et~al., \emph{{JADES: Differing assembly histories of galaxies
  -- Observational evidence for bursty SFHs and (mini-)quenching in the first
  billion years of the Universe}},
  \href{https://doi.org/10.48550/arXiv.2306.02470}{\emph{arXiv e-prints} (2023)
  arXiv:2306.02470} [\href{https://arxiv.org/abs/2306.02470}{{\ttfamily
  2306.02470}}].

\bibitem{Enders2023}
A.U.~{Enders}, D.J.~{Bomans} and A.~{Wittje}, \emph{{Lyman continuum leaker
  candidates among highly ionised, low-redshift dwarf galaxies selected from He
  II}}, \href{https://doi.org/10.1051/0004-6361/202245167}{\emph{\aap}
  {\bfseries 672} (2023) A11}
  [\href{https://arxiv.org/abs/2302.02868}{{\ttfamily 2302.02868}}].

\bibitem{Freese2009}
K.~{Freese}, P.~{Gondolo}, J.A.~{Sellwood} and D.~{Spolyar}, \emph{{Dark Matter
  Densities During the Formation of the First Stars and in Dark Stars}},
  \href{https://doi.org/10.1088/0004-637X/693/2/1563}{\emph{\apj} {\bfseries
  693} (2009) 1563} [\href{https://arxiv.org/abs/0805.3540}{{\ttfamily
  0805.3540}}].

\bibitem{Gnedin_2008}
N.Y.~Gnedin, A.V.~Kravtsov and H.-W.~Chen, \emph{Escape of ionizing radiation
  from high-redshift galaxies}, \href{https://doi.org/10.1086/524007}{\emph{The
  Astrophysical Journal} {\bfseries 672} (2008) 765}.

\bibitem{Calabro2024}
A.~{Calabr{\`o}}, L.~{Pentericci}, P.~{Santini}, A.~{Ferrara}, M.~{Llerena},
  S.~{Mascia} et~al., \emph{{The evolution of the SFR and $\Sigma_{SFR}$ of
  galaxies in cosmic morning ($4<z<10$)}},
  \href{https://doi.org/10.48550/arXiv.2402.17829}{\emph{arXiv e-prints} (2024)
  arXiv:2402.17829} [\href{https://arxiv.org/abs/2402.17829}{{\ttfamily
  2402.17829}}].

\bibitem{Ocvirk2016}
P.~{Ocvirk}, N.~{Gillet}, P.R.~{Shapiro}, D.~{Aubert}, I.T.~{Iliev},
  R.~{Teyssier} et~al., \emph{{Cosmic Dawn (CoDa): the First
  Radiation-Hydrodynamics Simulation of Reionization and Galaxy Formation in
  the Local Universe}},
  \href{https://doi.org/10.1093/mnras/stw2036}{\emph{\mnras} {\bfseries 463}
  (2016) 1462} [\href{https://arxiv.org/abs/1511.00011}{{\ttfamily
  1511.00011}}].

\bibitem{Lewis2020}
J.S.W.~{Lewis}, P.~{Ocvirk}, D.~{Aubert}, J.G.~{Sorce}, P.R.~{Shapiro},
  N.~{Deparis} et~al., \emph{{Galactic ionizing photon budget during the epoch
  of reionization in the Cosmic Dawn II simulation}},
  \href{https://doi.org/10.1093/mnras/staa1748}{\emph{\mnras} {\bfseries 496}
  (2020) 4342} [\href{https://arxiv.org/abs/2001.07785}{{\ttfamily
  2001.07785}}].

\bibitem{Mesinger2011}
A.~{Mesinger}, S.~{Furlanetto} and R.~{Cen}, \emph{{21CMFAST: a fast,
  seminumerical simulation of the high-redshift 21-cm signal}},
  \href{https://doi.org/10.1111/j.1365-2966.2010.17731.x}{\emph{\mnras}
  {\bfseries 411} (2011) 955}
  [\href{https://arxiv.org/abs/1003.3878}{{\ttfamily 1003.3878}}].

\bibitem{Ferrara2013}
A.~{Ferrara} and A.~{Loeb}, \emph{{Escape fraction of the ionizing radiation
  from starburst galaxies at high redshifts}},
  \href{https://doi.org/10.1093/mnras/stt381}{\emph{\mnras} {\bfseries 431}
  (2013) 2826} [\href{https://arxiv.org/abs/1209.2123}{{\ttfamily 1209.2123}}].

\bibitem{Wise_2009}
J.H.~Wise and R.~Cen, \emph{Ionizing photon escape fractions from high-redshift
  dwarf galaxies}, \href{https://doi.org/10.1088/0004-637X/693/1/984}{\emph{The
  Astrophysical Journal} {\bfseries 693} (2009) 984}.

\bibitem{Kostyuk2023}
I.~{Kostyuk}, B.~{Ciardi} and A.~{Ferrara}, \emph{{Physically motivated
  modeling of LyC escape fraction during reionization}},
  \href{https://doi.org/10.48550/arXiv.2308.01476}{\emph{arXiv e-prints} (2023)
  arXiv:2308.01476} [\href{https://arxiv.org/abs/2308.01476}{{\ttfamily
  2308.01476}}].

\bibitem{Lacey1985}
C.G.~{Lacey} and S.M.~{Fall}, \emph{{Chemical evolution of the galactic disk
  with radial gas flows.}}, \href{https://doi.org/10.1086/162970}{\emph{\apj}
  {\bfseries 290} (1985) 154}.

\bibitem{Matteucci1989}
F.~Matteucci and P.~François, \emph{{Galactic chemical evolution: abundance
  gradients of individual elements}},
  \href{https://doi.org/10.1093/mnras/239.3.885}{\emph{Monthly Notices of the
  Royal Astronomical Society} {\bfseries 239} (1989) 885}
  [\href{https://arxiv.org/abs/https://academic.oup.com/mnras/article-pdf/239/3/885/18194800/mnras239-0885.pdf}{{\ttfamily
  https://academic.oup.com/mnras/article-pdf/239/3/885/18194800/mnras239-0885.pdf}}].

\bibitem{White1991}
S.D.M.~{White} and C.S.~{Frenk}, \emph{{Galaxy Formation through Hierarchical
  Clustering}}, \href{https://doi.org/10.1086/170483}{\emph{\apj} {\bfseries
  379} (1991) 52}.

\bibitem{Japelj2017}
J.~Japelj, E.~Vanzella, F.~Fontanot, S.~Cristiani, G.B.~Caminha, P.~Tozzi
  et~al., \emph{{Constraints on the Lyman continuum escape fraction for faint
  star-forming galaxies}},
  \href{https://doi.org/10.1093/mnras/stx477}{\emph{Monthly Notices of the
  Royal Astronomical Society} {\bfseries 468} (2017) 389}
  [\href{https://arxiv.org/abs/https://academic.oup.com/mnras/article-pdf/468/1/389/11057672/stx477.pdf}{{\ttfamily
  https://academic.oup.com/mnras/article-pdf/468/1/389/11057672/stx477.pdf}}].

\bibitem{Leitherer1996}
C.~{Leitherer}, W.D.~{Vacca}, P.S.~{Conti}, A.V.~{Filippenko}, C.~{Robert} and
  W.L.W.~{Sargent}, \emph{{Hubble Space Telescope Ultraviolet Imaging and
  Spectroscopy of the Bright Starburst in the Wolf-Rayet Galaxy NGC 4214}},
  \href{https://doi.org/10.1086/177456}{\emph{\apj} {\bfseries 465} (1996)
  717}.

\bibitem{Heckman2011}
T.M.~{Heckman}, S.~{Borthakur}, R.~{Overzier}, G.~{Kauffmann}, A.~{Basu-Zych},
  C.~{Leitherer} et~al., \emph{{Extreme Feedback and the Epoch of Reionization:
  Clues in the Local Universe}},
  \href{https://doi.org/10.1088/0004-637X/730/1/5}{\emph{\apj} {\bfseries 730}
  (2011) 5} [\href{https://arxiv.org/abs/1101.4219}{{\ttfamily 1101.4219}}].

\bibitem{Shapley2006}
A.E.~{Shapley}, C.C.~{Steidel}, M.~{Pettini}, K.L.~{Adelberger} and D.K.~{Erb},
  \emph{{The Direct Detection of Lyman Continuum Emission from Star-forming
  Galaxies at z\raisebox{-0.5ex}\textasciitilde3}},
  \href{https://doi.org/10.1086/507511}{\emph{\apj} {\bfseries 651} (2006) 688}
  [\href{https://arxiv.org/abs/astro-ph/0606635}{{\ttfamily
  astro-ph/0606635}}].

\bibitem{Siana2007}
B.~{Siana}, H.I.~{Teplitz}, J.~{Colbert}, H.C.~{Ferguson}, M.~{Dickinson},
  T.M.~{Brown} et~al., \emph{{New Constraints on the Lyman Continuum Escape
  Fraction at z\raisebox{-0.5ex}\textasciitilde1.3}},
  \href{https://doi.org/10.1086/521185}{\emph{\apj} {\bfseries 668} (2007) 62}
  [\href{https://arxiv.org/abs/0706.4093}{{\ttfamily 0706.4093}}].

\bibitem{Vanzella_2010}
E.~{Vanzella}, M.~{Giavalisco}, A.K.~{Inoue}, M.~{Nonino}, F.~{Fontanot},
  S.~{Cristiani} et~al., \emph{{The Great Observatories Origins Deep Survey:
  Constraints on the Lyman Continuum Escape Fraction Distribution of
  Lyman-break Galaxies at 3.4 < z < 4.5}},
  \href{https://doi.org/10.1088/0004-637X/725/1/1011}{\emph{\apj} {\bfseries
  725} (2010) 1011} [\href{https://arxiv.org/abs/1009.1140}{{\ttfamily
  1009.1140}}].

\bibitem{Nestor_2011}
D.B.~Nestor, A.E.~Shapley, C.C.~Steidel and B.~Siana, \emph{Narrowband imaging
  of escaping lyman-continuum emission in the ssa22 field*,†},
  \href{https://doi.org/10.1088/0004-637X/736/1/18}{\emph{The Astrophysical
  Journal} {\bfseries 736} (2011) 18}.

\bibitem{Boutsia_2011}
K.~{Boutsia}, A.~{Grazian}, E.~{Giallongo}, A.~{Fontana}, L.~{Pentericci},
  M.~{Castellano} et~al., \emph{{A Low Escape Fraction of Ionizing Photons of L
  > L* Lyman Break Galaxies at z = 3.3}},
  \href{https://doi.org/10.1088/0004-637X/736/1/41}{\emph{\apj} {\bfseries 736}
  (2011) 41} [\href{https://arxiv.org/abs/1104.5237}{{\ttfamily 1104.5237}}].

\bibitem{Izotov2016}
Y.I.~{Izotov}, I.~{Orlitov{\'a}}, D.~{Schaerer}, T.X.~{Thuan}, A.~{Verhamme},
  N.G.~{Guseva} et~al., \emph{{Eight per cent leakage of Lyman continuum
  photons from a compact, star-forming dwarf galaxy}},
  \href{https://doi.org/10.1038/nature16456}{\emph{\nat} {\bfseries 529} (2016)
  178} [\href{https://arxiv.org/abs/1601.03068}{{\ttfamily 1601.03068}}].

\bibitem{Melinder2023}
J.~{Melinder}, G.~{{\"O}stlin}, M.~{Hayes}, A.~{Rasekh}, J.M.~{Mas-Hesse},
  J.M.~{Cannon} et~al., \emph{{The Ly{\ensuremath{\alpha}} Reference Sample.
  XIV. Ly{\ensuremath{\alpha}} Imaging of 45 Low-redshift Star-forming Galaxies
  and Inferences on Global Emission}},
  \href{https://doi.org/10.3847/1538-4365/acc2b8}{\emph{\apjs} {\bfseries 266}
  (2023) 15} [\href{https://arxiv.org/abs/2302.14077}{{\ttfamily 2302.14077}}].

\bibitem{Fernandez2011}
E.R.~{Fernandez} and J.M.~{Shull}, \emph{{The Effect of Galactic Properties on
  the Escape Fraction of Ionizing Photons}},
  \href{https://doi.org/10.1088/0004-637X/731/1/20}{\emph{\apj} {\bfseries 731}
  (2011) 20} [\href{https://arxiv.org/abs/1006.3519}{{\ttfamily 1006.3519}}].

\bibitem{Trebitsch2017}
M.~Trebitsch, J.~Blaizot, J.~Rosdahl, J.~Devriendt and A.~Slyz,
  \emph{{Fluctuating feedback-regulated escape fraction of ionizing radiation
  in low-mass, high-redshift galaxies}},
  \href{https://doi.org/10.1093/mnras/stx1060}{\emph{Monthly Notices of the
  Royal Astronomical Society} {\bfseries 470} (2017) 224}
  [\href{https://arxiv.org/abs/https://academic.oup.com/mnras/article-pdf/470/1/224/17843908/stx1060.pdf}{{\ttfamily
  https://academic.oup.com/mnras/article-pdf/470/1/224/17843908/stx1060.pdf}}].

\bibitem{Kimm2014}
T.~{Kimm} and R.~{Cen}, \emph{{Escape Fraction of Ionizing Photons during
  Reionization: Effects due to Supernova Feedback and Runaway OB Stars}},
  \href{https://doi.org/10.1088/0004-637X/788/2/121}{\emph{\apj} {\bfseries
  788} (2014) 121} [\href{https://arxiv.org/abs/1405.0552}{{\ttfamily
  1405.0552}}].

\bibitem{Wise2014}
J.H.~{Wise}, V.G.~{Demchenko}, M.T.~{Halicek}, M.L.~{Norman}, M.J.~{Turk},
  T.~{Abel} et~al., \emph{{The birth of a galaxy - III. Propelling reionization
  with the faintest galaxies}},
  \href{https://doi.org/10.1093/mnras/stu979}{\emph{\mnras} {\bfseries 442}
  (2014) 2560} [\href{https://arxiv.org/abs/1403.6123}{{\ttfamily 1403.6123}}].

\bibitem{Yajima2011}
H.~{Yajima}, J.-H.~{Choi} and K.~{Nagamine}, \emph{{Escape fraction of ionizing
  photons from high-redshift galaxies in cosmological SPH simulations}},
  \href{https://doi.org/10.1111/j.1365-2966.2010.17920.x}{\emph{\mnras}
  {\bfseries 412} (2011) 411}
  [\href{https://arxiv.org/abs/1002.3346}{{\ttfamily 1002.3346}}].

\bibitem{Schaye2015}
J.~{Schaye}, R.A.~{Crain}, R.G.~{Bower}, M.~{Furlong}, M.~{Schaller},
  T.~{Theuns} et~al., \emph{{The EAGLE project: simulating the evolution and
  assembly of galaxies and their environments}},
  \href{https://doi.org/10.1093/mnras/stu2058}{\emph{\mnras} {\bfseries 446}
  (2015) 521} [\href{https://arxiv.org/abs/1407.7040}{{\ttfamily 1407.7040}}].

\bibitem{Robertson2023a}
B.E.~{Robertson}, S.~{Tacchella}, B.D.~{Johnson}, K.~{Hainline}, L.~{Whitler},
  D.J.~{Eisenstein} et~al., \emph{{Identification and properties of intense
  star-forming galaxies at redshifts z > 10}},
  \href{https://doi.org/10.1038/s41550-023-01921-1}{\emph{Nature Astronomy}
  {\bfseries 7} (2023) 611} [\href{https://arxiv.org/abs/2212.04480}{{\ttfamily
  2212.04480}}].

\bibitem{robertson2023b}
B.~Robertson, B.D.~Johnson, S.~Tacchella, D.J.~Eisenstein, K.~Hainline,
  S.~Arribas et~al., \emph{Earliest galaxies in the jades origins field:
  Luminosity function and cosmic star-formation rate density 300 myr after the
  big bang},  2023.

\bibitem{McQuinn2023}
K.B.W.~{McQuinn}, M.J.B.~{Newman}, A.~{Savino}, A.E.~{Dolphin}, D.R.~{Weisz},
  B.F.~{Williams} et~al., \emph{{The JWST Resolved Stellar Populations Early
  Release Science Program IV: The Star Formation History of the Local Group
  Galaxy WLM}}, \href{https://doi.org/10.48550/arXiv.2312.03060}{\emph{arXiv
  e-prints} (2023) arXiv:2312.03060}
  [\href{https://arxiv.org/abs/2312.03060}{{\ttfamily 2312.03060}}].

\bibitem{Harikane_2023}
Y.~Harikane, M.~Ouchi, M.~Oguri, Y.~Ono, K.~Nakajima, Y.~Isobe et~al., \emph{A
  comprehensive study of galaxies at z ~ 9–16 found in the early jwst data:
  Ultraviolet luminosity functions and cosmic star formation history at the
  pre-reionization epoch},
  \href{https://doi.org/10.3847/1538-4365/acaaa9}{\emph{The Astrophysical
  Journal Supplement Series} {\bfseries 265} (2023) 5}.

\bibitem{Bouwens2023}
R.J.~{Bouwens}, M.~{Stefanon}, G.~{Brammer}, P.A.~{Oesch},
  T.~{Herard-Demanche}, G.D.~{Illingworth} et~al., \emph{{Evolution of the UV
  LF from z 15 to z 8 using new JWST NIRCam medium-band observations over the
  HUDF/XDF}}, \href{https://doi.org/10.1093/mnras/stad1145}{\emph{\mnras}
  {\bfseries 523} (2023) 1036}
  [\href{https://arxiv.org/abs/2211.02607}{{\ttfamily 2211.02607}}].

\bibitem{Faucher-Giguere2018}
C.-A.~{Faucher-Gigu{\`e}re}, \emph{{A model for the origin of bursty star
  formation in galaxies}},
  \href{https://doi.org/10.1093/mnras/stx2595}{\emph{\mnras} {\bfseries 473}
  (2018) 3717} [\href{https://arxiv.org/abs/1701.04824}{{\ttfamily
  1701.04824}}].

\bibitem{Borthakur2014}
S.~{Borthakur}, T.M.~{Heckman}, C.~{Leitherer} and R.A.~{Overzier}, \emph{{A
  local clue to the reionization of the universe}},
  \href{https://doi.org/10.1126/science.1254214}{\emph{Science} {\bfseries 346}
  (2014) 216} [\href{https://arxiv.org/abs/1410.3511}{{\ttfamily 1410.3511}}].

\bibitem{Schaerer2019}
D.~{Schaerer}, T.~{Fragos} and Y.I.~{Izotov}, \emph{{X-ray binaries as the
  origin of nebular He II emission in low-metallicity star-forming galaxies}},
  \href{https://doi.org/10.1051/0004-6361/201935005}{\emph{\aap} {\bfseries
  622} (2019) L10} [\href{https://arxiv.org/abs/1902.10496}{{\ttfamily
  1902.10496}}].

\bibitem{Park2019}
J.~{Park}, A.~{Mesinger}, B.~{Greig} and N.~{Gillet}, \emph{{Inferring the
  astrophysics of reionization and cosmic dawn from galaxy luminosity functions
  and the 21-cm signal}},
  \href{https://doi.org/10.1093/mnras/stz032}{\emph{\mnras} {\bfseries 484}
  (2019) 933} [\href{https://arxiv.org/abs/1809.08995}{{\ttfamily
  1809.08995}}].

\bibitem{Finkelstein2019}
S.L.~{Finkelstein}, A.~{D'Aloisio}, J.-P.~{Paardekooper}, J.~{Ryan}, Russell,
  P.~{Behroozi}, K.~{Finlator} et~al., \emph{{Conditions for Reionizing the
  Universe with a Low Galaxy Ionizing Photon Escape Fraction}},
  \href{https://doi.org/10.3847/1538-4357/ab1ea8}{\emph{\apj} {\bfseries 879}
  (2019) 36} [\href{https://arxiv.org/abs/1902.02792}{{\ttfamily 1902.02792}}].

\bibitem{Marinacci2013}
F.~Marinacci, R.~Pakmor and V.~Springel, \emph{{The formation of disc galaxies
  in high-resolution moving-mesh cosmological simulations}},
  \href{https://doi.org/10.1093/mnras/stt2003}{\emph{Monthly Notices of the
  Royal Astronomical Society} {\bfseries 437} (2013) 1750}
  [\href{https://arxiv.org/abs/https://academic.oup.com/mnras/article-pdf/437/2/1750/3877129/stt2003.pdf}{{\ttfamily
  https://academic.oup.com/mnras/article-pdf/437/2/1750/3877129/stt2003.pdf}}].

\bibitem{Speagle2014}
J.S.~{Speagle}, C.L.~{Steinhardt}, P.L.~{Capak} and J.D.~{Silverman}, \emph{{A
  Highly Consistent Framework for the Evolution of the Star-Forming ``Main
  Sequence'' from z \raisebox{-0.5ex}\textasciitilde 0-6}},
  \href{https://doi.org/10.1088/0067-0049/214/2/15}{\emph{\apjs} {\bfseries
  214} (2014) 15} [\href{https://arxiv.org/abs/1405.2041}{{\ttfamily
  1405.2041}}].

\bibitem{Popesso2023}
P.~{Popesso}, A.~{Concas}, G.~{Cresci}, S.~{Belli}, G.~{Rodighiero}, H.~{Inami}
  et~al., \emph{{The main sequence of star-forming galaxies across cosmic
  times}}, \href{https://doi.org/10.1093/mnras/stac3214}{\emph{\mnras}
  {\bfseries 519} (2023) 1526}
  [\href{https://arxiv.org/abs/2203.10487}{{\ttfamily 2203.10487}}].

\bibitem{Tacchella2020}
S.~{Tacchella}, J.C.~{Forbes} and N.~{Caplar}, \emph{{Stochastic modelling of
  star-formation histories II: star-formation variability from molecular clouds
  and gas inflow}}, \href{https://doi.org/10.1093/mnras/staa1838}{\emph{\mnras}
  {\bfseries 497} (2020) 698}
  [\href{https://arxiv.org/abs/2006.09382}{{\ttfamily 2006.09382}}].

\bibitem{Shuntov2022}
M.~{Shuntov}, H.J.~{McCracken}, R.~{Gavazzi}, C.~{Laigle}, J.R.~{Weaver},
  I.~{Davidzon} et~al., \emph{{COSMOS2020: Cosmic evolution of the
  stellar-to-halo mass relation for central and satellite galaxies up to z
  {\ensuremath{\sim}} 5}},
  \href{https://doi.org/10.1051/0004-6361/202243136}{\emph{\aap} {\bfseries
  664} (2022) A61} [\href{https://arxiv.org/abs/2203.10895}{{\ttfamily
  2203.10895}}].

\bibitem{Golden2022}
J.B.~{Golden-Marx}, C.J.~{Miller}, Y.~{Zhang}, R.L.C.~{Ogando}, A.~{Palmese},
  T.M.C.~{Abbott} et~al., \emph{{The Observed Evolution of the Stellar
  Mass-Halo Mass Relation for Brightest Central Galaxies}},
  \href{https://doi.org/10.3847/1538-4357/ac4cb4}{\emph{\apj} {\bfseries 928}
  (2022) 28} [\href{https://arxiv.org/abs/2107.02197}{{\ttfamily 2107.02197}}].

\bibitem{Matthee2017}
J.~{Matthee}, J.~{Schaye}, R.A.~{Crain}, M.~{Schaller}, R.~{Bower} and
  T.~{Theuns}, \emph{{The origin of scatter in the stellar mass-halo mass
  relation of central galaxies in the EAGLE simulation}},
  \href{https://doi.org/10.1093/mnras/stw2884}{\emph{\mnras} {\bfseries 465}
  (2017) 2381} [\href{https://arxiv.org/abs/1608.08218}{{\ttfamily
  1608.08218}}].

\bibitem{Kimm2017}
T.~Kimm, H.~Katz, M.~Haehnelt, J.~Rosdahl, J.~Devriendt and A.~Slyz,
  \emph{{Feedback-regulated star formation and escape of LyC photons from
  mini-haloes during reionization}},
  \href{https://doi.org/10.1093/mnras/stx052}{\emph{Monthly Notices of the
  Royal Astronomical Society} {\bfseries 466} (2017) 4826}
  [\href{https://arxiv.org/abs/https://academic.oup.com/mnras/article-pdf/466/4/4826/10873852/stx052.pdf}{{\ttfamily
  https://academic.oup.com/mnras/article-pdf/466/4/4826/10873852/stx052.pdf}}].

\bibitem{Tachella16}
S.~Tacchella, A.~Dekel, C.M.~Carollo, D.~Ceverino, C.~DeGraf, S.~Lapiner
  et~al., \emph{{The confinement of star-forming galaxies into a main sequence
  through episodes of gas compaction, depletion and replenishment}},
  \href{https://doi.org/10.1093/mnras/stw131}{\emph{Monthly Notices of the
  Royal Astronomical Society} {\bfseries 457} (2016) 2790}
  [\href{https://arxiv.org/abs/https://academic.oup.com/mnras/article-pdf/457/3/2790/8000909/stw131.pdf}{{\ttfamily
  https://academic.oup.com/mnras/article-pdf/457/3/2790/8000909/stw131.pdf}}].

\bibitem{Sharma2018CEMP}
M.~{Sharma}, T.~{Theuns}, C.S.~{Frenk} and R.J.~{Cooke}, \emph{{Origins of
  carbon-enhanced metal-poor stars}},
  \href{https://doi.org/10.1093/mnras/stx2392}{\emph{\mnras} {\bfseries 473}
  (2018) 984} [\href{https://arxiv.org/abs/1611.03868}{{\ttfamily
  1611.03868}}].

\bibitem{Sharma2019}
M.~{Sharma}, T.~{Theuns} and C.~{Frenk}, \emph{{The chemical imprint of the
  bursty nature of Milky Way's progenitors}},
  \href{https://doi.org/10.1093/mnrasl/sly195}{\emph{\mnras} {\bfseries 482}
  (2019) L145} [\href{https://arxiv.org/abs/1805.05342}{{\ttfamily
  1805.05342}}].

\bibitem{Sun2023}
G.~{Sun}, C.-A.~{Faucher-Gigu{\`e}re}, C.C.~{Hayward} and X.~{Shen},
  \emph{{Seen and unseen: bursty star formation and its implications for
  observations of high-redshift galaxies with JWST}},
  \href{https://doi.org/10.1093/mnras/stad2902}{\emph{\mnras} {\bfseries 526}
  (2023) 2665} [\href{https://arxiv.org/abs/2305.02713}{{\ttfamily
  2305.02713}}].

\bibitem{Nikolic2024}
I.~{Nikoli{\'c}}, A.~{Mesinger}, J.E.~{Davies} and D.~{Prelogovi{\'c}},
  \emph{{The importance of stochasticity in determining galaxy emissivities and
  UV LFs during cosmic dawn and reionization}},
  \href{https://doi.org/10.48550/arXiv.2406.15237}{\emph{arXiv e-prints} (2024)
  arXiv:2406.15237} [\href{https://arxiv.org/abs/2406.15237}{{\ttfamily
  2406.15237}}].

\bibitem{Bunker2023}
A.J.~{Bunker}, A.J.~{Cameron}, E.~{Curtis-Lake}, P.~{Jakobsen}, S.~{Carniani},
  M.~{Curti} et~al., \emph{{JADES NIRSpec Initial Data Release for the Hubble
  Ultra Deep Field: Redshifts and Line Fluxes of Distant Galaxies from the
  Deepest JWST Cycle 1 NIRSpec Multi-Object Spectroscopy}},
  \href{https://doi.org/10.48550/arXiv.2306.02467}{\emph{arXiv e-prints} (2023)
  arXiv:2306.02467} [\href{https://arxiv.org/abs/2306.02467}{{\ttfamily
  2306.02467}}].

\bibitem{Jones2023}
G.C.~{Jones}, A.J.~{Bunker}, A.~{Saxena}, J.~{Witstok}, D.P.~{Stark},
  S.~{Arribas} et~al., \emph{{JADES: The emergence and evolution of Ly$\alpha$
  emission and constraints on the IGM neutral fraction}},
  \href{https://doi.org/10.48550/arXiv.2306.02471}{\emph{arXiv e-prints} (2023)
  arXiv:2306.02471} [\href{https://arxiv.org/abs/2306.02471}{{\ttfamily
  2306.02471}}].

\bibitem{Curti2023}
M.~{Curti}, R.~{Maiolino}, E.~{Curtis-Lake}, J.~{Chevallard}, S.~{Carniani},
  F.~{D'Eugenio} et~al., \emph{{JADES: Insights on the low-mass end of the
  mass--metallicity--star-formation rate relation at $3 < z < 10$ from deep
  JWST/NIRSpec spectroscopy}},
  \href{https://doi.org/10.48550/arXiv.2304.08516}{\emph{arXiv e-prints} (2023)
  arXiv:2304.08516} [\href{https://arxiv.org/abs/2304.08516}{{\ttfamily
  2304.08516}}].

\bibitem{Morales2023}
A.M.~{Morales}, S.L.~{Finkelstein}, G.C.K.~{Leung}, M.B.~{Bagley},
  N.J.~{Cleri}, R.~{Dave} et~al., \emph{{Rest-Frame UV Colors for Faint
  Galaxies at $z \sim 9-16$ with the JWST NGDEEP Survey}},
  \href{https://doi.org/10.48550/arXiv.2311.04294}{\emph{arXiv e-prints} (2023)
  arXiv:2311.04294} [\href{https://arxiv.org/abs/2311.04294}{{\ttfamily
  2311.04294}}].

\bibitem{Nakajima2023}
K.~{Nakajima}, M.~{Ouchi}, Y.~{Isobe}, Y.~{Harikane}, Y.~{Zhang}, Y.~{Ono}
  et~al., \emph{{JWST Census for the Mass-Metallicity Star Formation Relations
  at z = 4-10 with Self-consistent Flux Calibration and Proper Metallicity
  Calibrators}}, \href{https://doi.org/10.3847/1538-4365/acd556}{\emph{\apjs}
  {\bfseries 269} (2023) 33}
  [\href{https://arxiv.org/abs/2301.12825}{{\ttfamily 2301.12825}}].

\bibitem{Jung2023}
I.~{Jung}, S.L.~{Finkelstein}, P.~{Arrabal Haro}, M.~{Dickinson},
  H.C.~{Ferguson}, T.A.~{Hutchison} et~al., \emph{{CEERS: Diversity of
  Lyman-Alpha Emitters during the Epoch of Reionization}},
  \href{https://doi.org/10.48550/arXiv.2304.05385}{\emph{arXiv e-prints} (2023)
  arXiv:2304.05385} [\href{https://arxiv.org/abs/2304.05385}{{\ttfamily
  2304.05385}}].

\bibitem{Haro2023}
P.~{Arrabal Haro}, M.~{Dickinson}, S.L.~{Finkelstein}, J.S.~{Kartaltepe},
  C.T.~{Donnan}, D.~{Burgarella} et~al., \emph{{Confirmation and refutation of
  very luminous galaxies in the early Universe}},
  \href{https://doi.org/10.1038/s41586-023-06521-7}{\emph{\nat} {\bfseries 622}
  (2023) 707} [\href{https://arxiv.org/abs/2303.15431}{{\ttfamily
  2303.15431}}].

\bibitem{Harikane2024}
Y.~{Harikane}, Y.~{Zhang}, K.~{Nakajima}, M.~{Ouchi}, Y.~{Isobe}, Y.~{Ono}
  et~al., \emph{{A JWST/NIRSpec First Census of Broad-line AGNs at z = 4-7:
  Detection of 10 Faint AGNs with M $_{BH}$ {}10$^{6}$-{}10$^{8}$ M
  $_{{\ensuremath{\odot}}}$ and Their Host Galaxy Properties}},
  \href{https://doi.org/10.3847/1538-4357/ad029e}{\emph{\apj} {\bfseries 959}
  (2023) 39} [\href{https://arxiv.org/abs/2303.11946}{{\ttfamily 2303.11946}}].

\bibitem{Zitrin2015}
A.~{Zitrin}, I.~{Labb{\'e}}, S.~{Belli}, R.~{Bouwens}, R.S.~{Ellis},
  G.~{Roberts-Borsani} et~al., \emph{{Lyman{\ensuremath{\alpha}} Emission from
  a Luminous z = 8.68 Galaxy: Implications for Galaxies as Tracers of Cosmic
  Reionization}},
  \href{https://doi.org/10.1088/2041-8205/810/1/L12}{\emph{\apjl} {\bfseries
  810} (2015) L12} [\href{https://arxiv.org/abs/1507.02679}{{\ttfamily
  1507.02679}}].

\bibitem{Fujimoto2023}
S.~{Fujimoto}, P.~{Arrabal Haro}, M.~{Dickinson}, S.L.~{Finkelstein},
  J.S.~{Kartaltepe}, R.L.~{Larson} et~al., \emph{{CEERS Spectroscopic
  Confirmation of NIRCam-selected z {\ensuremath{\gtrsim}} 8 Galaxy Candidates
  with JWST/NIRSpec: Initial Characterization of Their Properties}},
  \href{https://doi.org/10.3847/2041-8213/acd2d9}{\emph{\apjl} {\bfseries 949}
  (2023) L25} [\href{https://arxiv.org/abs/2301.09482}{{\ttfamily
  2301.09482}}].

\bibitem{Tang2023}
M.~Tang, D.P.~Stark, Z.~Chen, C.~Mason, M.~Topping, R.~Endsley et~al.,
  \emph{{JWST/NIRSpec spectroscopy of z = 7–9 star-forming galaxies with
  CEERS: new insight into bright Ly-$\alpha$ emitters in ionized bubbles}},
  \href{https://doi.org/10.1093/mnras/stad2763}{\emph{Monthly Notices of the
  Royal Astronomical Society} {\bfseries 526} (2023) 1657}
  [\href{https://arxiv.org/abs/https://academic.oup.com/mnras/article-pdf/526/2/1657/51799799/stad2763.pdf}{{\ttfamily
  https://academic.oup.com/mnras/article-pdf/526/2/1657/51799799/stad2763.pdf}}].

\bibitem{Larson2023}
R.L.~{Larson}, S.L.~{Finkelstein}, D.D.~{Kocevski}, T.A.~{Hutchison},
  J.R.~{Trump}, P.~{Arrabal Haro} et~al., \emph{{A CEERS Discovery of an
  Accreting Supermassive Black Hole 570 Myr after the Big Bang: Identifying a
  Progenitor of Massive z > 6 Quasars}},
  \href{https://doi.org/10.3847/2041-8213/ace619}{\emph{\apjl} {\bfseries 953}
  (2023) L29} [\href{https://arxiv.org/abs/2303.08918}{{\ttfamily
  2303.08918}}].

\bibitem{Sanders2023}
R.L.~{Sanders}, A.E.~{Shapley}, M.W.~{Topping}, N.A.~{Reddy} and
  G.B.~{Brammer}, \emph{{Excitation and Ionization Properties of Star-forming
  Galaxies at z = 2.0-9.3 with JWST/NIRSpec}},
  \href{https://doi.org/10.3847/1538-4357/acedad}{\emph{\apj} {\bfseries 955}
  (2023) 54} [\href{https://arxiv.org/abs/2301.06696}{{\ttfamily 2301.06696}}].

\bibitem{Bunker2023b}
A.J.~{Bunker}, A.~{Saxena}, A.J.~{Cameron}, C.J.~{Willott}, E.~{Curtis-Lake},
  P.~{Jakobsen} et~al., \emph{{JADES NIRSpec Spectroscopy of GN-z11:
  Lyman-{\ensuremath{\alpha}} emission and possible enhanced nitrogen abundance
  in a z = 10.60 luminous galaxy}},
  \href{https://doi.org/10.1051/0004-6361/202346159}{\emph{\aap} {\bfseries
  677} (2023) A88} [\href{https://arxiv.org/abs/2302.07256}{{\ttfamily
  2302.07256}}].

\bibitem{Springel2005}
V.~{Springel}, S.D.M.~{White}, A.~{Jenkins}, C.S.~{Frenk}, N.~{Yoshida},
  L.~{Gao} et~al., \emph{{Simulations of the formation, evolution and
  clustering of galaxies and quasars}},
  \href{https://doi.org/10.1038/nature03597}{\emph{\nat} {\bfseries 435} (2005)
  629} [\href{https://arxiv.org/abs/astro-ph/0504097}{{\ttfamily
  astro-ph/0504097}}].

\bibitem{Vogelsberger2014}
M.~{Vogelsberger}, S.~{Genel}, V.~{Springel}, P.~{Torrey}, D.~{Sijacki},
  D.~{Xu} et~al., \emph{{Properties of galaxies reproduced by a hydrodynamic
  simulation}}, \href{https://doi.org/10.1038/nature13316}{\emph{\nat}
  {\bfseries 509} (2014) 177}
  [\href{https://arxiv.org/abs/1405.1418}{{\ttfamily 1405.1418}}].

\bibitem{Schaller2015}
M.~{Schaller}, C.~{Dalla Vecchia}, J.~{Schaye}, R.G.~{Bower}, T.~{Theuns},
  R.A.~{Crain} et~al., \emph{{The EAGLE simulations of galaxy formation: the
  importance of the hydrodynamics scheme}},
  \href{https://doi.org/10.1093/mnras/stv2169}{\emph{\mnras} {\bfseries 454}
  (2015) 2277} [\href{https://arxiv.org/abs/1509.05056}{{\ttfamily
  1509.05056}}].

\bibitem{Crain2015}
R.A.~{Crain}, J.~{Schaye}, R.G.~{Bower}, M.~{Furlong}, M.~{Schaller},
  T.~{Theuns} et~al., \emph{{The EAGLE simulations of galaxy formation:
  calibration of subgrid physics and model variations}},
  \href{https://doi.org/10.1093/mnras/stv725}{\emph{\mnras} {\bfseries 450}
  (2015) 1937} [\href{https://arxiv.org/abs/1501.01311}{{\ttfamily
  1501.01311}}].

\bibitem{DallaVecchia2008}
C.~{Dalla Vecchia} and J.~{Schaye}, \emph{{Simulating galactic outflows with
  kinetic supernova feedback}},
  \href{https://doi.org/10.1111/j.1365-2966.2008.13322.x}{\emph{\mnras}
  {\bfseries 387} (2008) 1431}
  [\href{https://arxiv.org/abs/0801.2770}{{\ttfamily 0801.2770}}].

\bibitem{Schaye2010}
J.~{Schaye}, C.~{Dalla Vecchia}, C.M.~{Booth}, R.P.C.~{Wiersma}, T.~{Theuns},
  M.R.~{Haas} et~al., \emph{{The physics driving the cosmic star formation
  history}},
  \href{https://doi.org/10.1111/j.1365-2966.2009.16029.x}{\emph{\mnras}
  {\bfseries 402} (2010) 1536}
  [\href{https://arxiv.org/abs/0909.5196}{{\ttfamily 0909.5196}}].

\bibitem{Kennicutt1998}
J.~{Kennicutt}, Robert~C., \emph{{The Global Schmidt Law in Star-forming
  Galaxies}}, \href{https://doi.org/10.1086/305588}{\emph{\apj} {\bfseries 498}
  (1998) 541} [\href{https://arxiv.org/abs/astro-ph/9712213}{{\ttfamily
  astro-ph/9712213}}].

\bibitem{Schaye2008}
J.~{Schaye} and C.~{Dalla Vecchia}, \emph{{On the relation between the Schmidt
  and Kennicutt-Schmidt star formation laws and its implications for numerical
  simulations}},
  \href{https://doi.org/10.1111/j.1365-2966.2007.12639.x}{\emph{\mnras}
  {\bfseries 383} (2008) 1210}
  [\href{https://arxiv.org/abs/0709.0292}{{\ttfamily 0709.0292}}].

\bibitem{Sparre2017}
M.~Sparre, C.C.~Hayward, R.~Feldmann, C.-A.~Faucher-Giguère, A.L.~Muratov,
  D.~Kereš et~al., \emph{{(Star)bursts of FIRE: observational signatures of
  bursty star formation in galaxies}},
  \href{https://doi.org/10.1093/mnras/stw3011}{\emph{Monthly Notices of the
  Royal Astronomical Society} {\bfseries 466} (2016) 88}
  [\href{https://arxiv.org/abs/https://academic.oup.com/mnras/article-pdf/466/1/88/10865056/stw3011.pdf}{{\ttfamily
  https://academic.oup.com/mnras/article-pdf/466/1/88/10865056/stw3011.pdf}}].

\bibitem{Zaroubi2013}
S.~{Zaroubi}, \emph{{The Epoch of Reionization}},  in \emph{The First
  Galaxies}, T.~{Wiklind}, B.~{Mobasher} and V.~{Bromm}, eds., vol.~396 of
  \emph{Astrophysics and Space Science Library}, p.~45, Jan., 2013,
  \href{https://doi.org/10.1007/978-3-642-32362-1_2}{DOI}
  [\href{https://arxiv.org/abs/1206.0267}{{\ttfamily 1206.0267}}].

\bibitem{Press1974}
W.H.~{Press} and P.~{Schechter}, \emph{{Formation of Galaxies and Clusters of
  Galaxies by Self-Similar Gravitational Condensation}},
  \href{https://doi.org/10.1086/152650}{\emph{\apj} {\bfseries 187} (1974)
  425}.

\bibitem{Mesinger2007}
A.~{Mesinger} and S.~{Furlanetto}, \emph{{Efficient Simulations of Early
  Structure Formation and Reionization}},
  \href{https://doi.org/10.1086/521806}{\emph{\apj} {\bfseries 669} (2007) 663}
  [\href{https://arxiv.org/abs/0704.0946}{{\ttfamily 0704.0946}}].

\bibitem{Mutch2016}
S.J.~{Mutch}, P.M.~{Geil}, G.B.~{Poole}, P.W.~{Angel}, A.R.~{Duffy},
  A.~{Mesinger} et~al., \emph{{Dark-ages reionization and galaxy formation
  simulation - III. Modelling galaxy formation and the epoch of reionization}},
  \href{https://doi.org/10.1093/mnras/stw1506}{\emph{\mnras} {\bfseries 462}
  (2016) 250} [\href{https://arxiv.org/abs/1512.00562}{{\ttfamily
  1512.00562}}].

\bibitem{Garaldi2022}
E.~{Garaldi}, R.~{Kannan}, A.~{Smith}, V.~{Springel}, R.~{Pakmor},
  M.~{Vogelsberger} et~al., \emph{{The THESAN project: properties of the
  intergalactic medium and its connection to reionization-era galaxies}},
  \href{https://doi.org/10.1093/mnras/stac257}{\emph{\mnras} {\bfseries 512}
  (2022) 4909} [\href{https://arxiv.org/abs/2110.01628}{{\ttfamily
  2110.01628}}].

\bibitem{Harrison1970}
E.R.~Harrison, \emph{Fluctuations at the threshold of classical cosmology},
  \href{https://doi.org/10.1103/PhysRevD.1.2726}{\emph{Phys. Rev. D} {\bfseries
  1} (1970) 2726}.

\bibitem{Zeldovich1972}
Y.B.~Zeldovich, \emph{{A Hypothesis, Unifying the Structure and the Entropy of
  the Universe}}, \href{https://doi.org/10.1093/mnras/160.1.1P}{\emph{Monthly
  Notices of the Royal Astronomical Society} {\bfseries 160} (1972) 1P}
  [\href{https://arxiv.org/abs/https://academic.oup.com/mnras/article-pdf/160/1/1P/8079415/mnras160-001P.pdf}{{\ttfamily
  https://academic.oup.com/mnras/article-pdf/160/1/1P/8079415/mnras160-001P.pdf}}].

\bibitem{Greig2015}
B.~{Greig} and A.~{Mesinger}, \emph{{21CMMC: an MCMC analysis tool enabling
  astrophysical parameter studies of the cosmic 21 cm signal}},
  \href{https://doi.org/10.1093/mnras/stv571}{\emph{\mnras} {\bfseries 449}
  (2015) 4246} [\href{https://arxiv.org/abs/1501.06576}{{\ttfamily
  1501.06576}}].

\bibitem{Greig2017}
B.~{Greig} and A.~{Mesinger}, \emph{{The global history of reionization}},
  \href{https://doi.org/10.1093/mnras/stw3026}{\emph{\mnras} {\bfseries 465}
  (2017) 4838} [\href{https://arxiv.org/abs/1605.05374}{{\ttfamily
  1605.05374}}].

\bibitem{Greig2018}
B.~{Greig} and A.~{Mesinger}, \emph{{21CMMC with a 3D light-cone: the impact of
  the co-evolution approximation on the astrophysics of reionization and cosmic
  dawn}}, \href{https://doi.org/10.1093/mnras/sty796}{\emph{\mnras} {\bfseries
  477} (2018) 3217} [\href{https://arxiv.org/abs/1801.01592}{{\ttfamily
  1801.01592}}].

\bibitem{Ma2015}
X.~Ma, D.~Kasen, P.F.~Hopkins, C.-A.~Faucher-Giguère, E.~Quataert, D.~Kereš
  et~al., \emph{{ The difficulty of getting high escape fractions of ionizing
  photons from high-redshift galaxies: a view from the FIRE cosmological
  simulations}}, \href{https://doi.org/10.1093/mnras/stv1679}{\emph{Monthly
  Notices of the Royal Astronomical Society} {\bfseries 453} (2015) 960}
  [\href{https://arxiv.org/abs/https://academic.oup.com/mnras/article-pdf/453/1/960/4935561/stv1679.pdf}{{\ttfamily
  https://academic.oup.com/mnras/article-pdf/453/1/960/4935561/stv1679.pdf}}].

\bibitem{Bouwens2019}
R.J.~{Bouwens}, M.~{Stefanon}, P.A.~{Oesch}, G.D.~{Illingworth},
  T.~{Nanayakkara}, G.~{Roberts-Borsani} et~al., \emph{{Newly Discovered Bright
  z {\ensuremath{\sim}} 9-10 Galaxies and Improved Constraints on Their
  Prevalence Using the Full CANDELS Area}},
  \href{https://doi.org/10.3847/1538-4357/ab24c5}{\emph{\apj} {\bfseries 880}
  (2019) 25} [\href{https://arxiv.org/abs/1905.05202}{{\ttfamily 1905.05202}}].

\bibitem{Naidu2022}
R.P.~{Naidu}, J.~{Matthee}, P.A.~{Oesch}, C.~{Conroy}, D.~{Sobral},
  G.~{Pezzulli} et~al., \emph{{The synchrony of production and escape: half the
  bright Ly{\ensuremath{\alpha}} emitters at z {\ensuremath{\approx}} 2 have
  Lyman continuum escape fractions {\ensuremath{\approx}}50 per cent}},
  \href{https://doi.org/10.1093/mnras/stab3601}{\emph{\mnras} {\bfseries 510}
  (2022) 4582} [\href{https://arxiv.org/abs/2110.11961}{{\ttfamily
  2110.11961}}].

\bibitem{Naidu2020}
R.P.~{Naidu}, S.~{Tacchella}, C.A.~{Mason}, S.~{Bose}, P.A.~{Oesch} and
  C.~{Conroy}, \emph{{Rapid Reionization by the Oligarchs: The Case for
  Massive, UV-bright, Star-forming Galaxies with High Escape Fractions}},
  \href{https://doi.org/10.3847/1538-4357/ab7cc9}{\emph{\apj} {\bfseries 892}
  (2020) 109} [\href{https://arxiv.org/abs/1907.13130}{{\ttfamily
  1907.13130}}].

\bibitem{Ouchi2018}
M.~{Ouchi}, Y.~{Harikane}, T.~{Shibuya}, K.~{Shimasaku}, Y.~{Taniguchi},
  A.~{Konno} et~al., \emph{{Systematic Identification of LAEs for Visible
  Exploration and Reionization Research Using Subaru HSC (SILVERRUSH). I.
  Program strategy and clustering properties of {\ensuremath{\sim}}2000
  Ly{\ensuremath{\alpha}} emitters at z = 6-7 over the 0.3-0.5 Gpc$^{2}$ survey
  area}}, \href{https://doi.org/10.1093/pasj/psx074}{\emph{\pasj} {\bfseries
  70} (2018) S13} [\href{https://arxiv.org/abs/1704.07455}{{\ttfamily
  1704.07455}}].

\bibitem{Ouchi2010}
M.~{Ouchi}, K.~{Shimasaku}, H.~{Furusawa}, T.~{Saito}, M.~{Yoshida},
  M.~{Akiyama} et~al., \emph{{Statistics of 207 Ly{\ensuremath{\alpha}}
  Emitters at a Redshift Near 7: Constraints on Reionization and Galaxy
  Formation Models}},
  \href{https://doi.org/10.1088/0004-637X/723/1/869}{\emph{\apj} {\bfseries
  723} (2010) 869} [\href{https://arxiv.org/abs/1007.2961}{{\ttfamily
  1007.2961}}].

\bibitem{Inoue2018}
A.K.~{Inoue}, K.~{Hasegawa}, T.~{Ishiyama}, H.~{Yajima}, I.~{Shimizu},
  M.~{Umemura} et~al., \emph{{SILVERRUSH. VI. A simulation of
  Ly{\ensuremath{\alpha}} emitters in the reionization epoch and a comparison
  with Subaru Hyper Suprime-Cam survey early data}},
  \href{https://doi.org/10.1093/pasj/psy048}{\emph{\pasj} {\bfseries 70} (2018)
  55} [\href{https://arxiv.org/abs/1801.00067}{{\ttfamily 1801.00067}}].

\bibitem{Morales2021}
A.M.~{Morales}, C.A.~{Mason}, S.~{Bruton}, M.~{Gronke}, F.~{Haardt} and
  C.~{Scarlata}, \emph{{The Evolution of the Lyman-alpha Luminosity Function
  during Reionization}},
  \href{https://doi.org/10.3847/1538-4357/ac1104}{\emph{\apj} {\bfseries 919}
  (2021) 120} [\href{https://arxiv.org/abs/2101.01205}{{\ttfamily
  2101.01205}}].

\bibitem{Itoh2018}
R.~{Itoh}, M.~{Ouchi}, H.~{Zhang}, A.K.~{Inoue}, K.~{Mawatari}, T.~{Shibuya}
  et~al., \emph{{CHORUS. II. Subaru/HSC Determination of the
  Ly{\ensuremath{\alpha}} Luminosity Function at z = 7.0: Constraints on Cosmic
  Reionization Model Parameter}},
  \href{https://doi.org/10.3847/1538-4357/aadfe4}{\emph{\apj} {\bfseries 867}
  (2018) 46} [\href{https://arxiv.org/abs/1805.05944}{{\ttfamily 1805.05944}}].

\bibitem{Goto2021}
H.~{Goto}, K.~{Shimasaku}, S.~{Yamanaka}, R.~{Momose}, M.~{Ando}, Y.~{Harikane}
  et~al., \emph{{SILVERRUSH. XI. Constraints on the Ly{\ensuremath{\alpha}}
  Luminosity Function and Cosmic Reionization at z = 7.3 with Subaru/Hyper
  Suprime-Cam}}, \href{https://doi.org/10.3847/1538-4357/ac308b}{\emph{\apj}
  {\bfseries 923} (2021) 229}
  [\href{https://arxiv.org/abs/2110.14474}{{\ttfamily 2110.14474}}].

\bibitem{Curtis-Lake2023}
E.~{Curtis-Lake}, S.~{Carniani}, A.~{Cameron}, S.~{Charlot}, P.~{Jakobsen},
  R.~{Maiolino} et~al., \emph{{Spectroscopic confirmation of four metal-poor
  galaxies at z = 10.3-13.2}},
  \href{https://doi.org/10.1038/s41550-023-01918-w}{\emph{Nature Astronomy}
  {\bfseries 7} (2023) 622} [\href{https://arxiv.org/abs/2212.04568}{{\ttfamily
  2212.04568}}].

\bibitem{Hsiao2023}
T.Y.-Y.~{Hsiao}, {Abdurro'uf}, D.~{Coe}, R.L.~{Larson}, I.~{Jung},
  M.~{Mingozzi} et~al., \emph{{JWST NIRSpec spectroscopy of the triply-lensed
  $z = 10.17$ galaxy MACS0647$-$JD}},
  \href{https://doi.org/10.48550/arXiv.2305.03042}{\emph{arXiv e-prints} (2023)
  arXiv:2305.03042} [\href{https://arxiv.org/abs/2305.03042}{{\ttfamily
  2305.03042}}].

\bibitem{Hoag2019}
A.~{Hoag}, M.~{Brada{\v{c}}}, K.~{Huang}, C.~{Mason}, T.~{Treu}, K.B.~{Schmidt}
  et~al., \emph{{Constraining the Neutral Fraction of Hydrogen in the IGM at
  Redshift 7.5}}, \href{https://doi.org/10.3847/1538-4357/ab1de7}{\emph{\apj}
  {\bfseries 878} (2019) 12}
  [\href{https://arxiv.org/abs/1901.09001}{{\ttfamily 1901.09001}}].

\bibitem{Mason2019}
C.A.~{Mason}, A.~{Fontana}, T.~{Treu}, K.B.~{Schmidt}, A.~{Hoag}, L.~{Abramson}
  et~al., \emph{{Inferences on the timeline of reionization at z
  {\ensuremath{\sim}} 8 from the KMOS Lens-Amplified Spectroscopic Survey}},
  \href{https://doi.org/10.1093/mnras/stz632}{\emph{\mnras} {\bfseries 485}
  (2019) 3947} [\href{https://arxiv.org/abs/1901.11045}{{\ttfamily
  1901.11045}}].

\bibitem{Jung2020}
I.~{Jung}, S.L.~{Finkelstein}, M.~{Dickinson}, T.A.~{Hutchison}, R.L.~{Larson},
  C.~{Papovich} et~al., \emph{{Texas Spectroscopic Search for
  Ly{\ensuremath{\alpha}} Emission at the End of Reionization. III. The
  Ly{\ensuremath{\alpha}} Equivalent-width Distribution and Ionized Structures
  at z > 7}}, \href{https://doi.org/10.3847/1538-4357/abbd44}{\emph{\apj}
  {\bfseries 904} (2020) 144}
  [\href{https://arxiv.org/abs/2009.10092}{{\ttfamily 2009.10092}}].

\bibitem{Whitler2020}
L.R.~{Whitler}, C.A.~{Mason}, K.~{Ren}, M.~{Dijkstra}, A.~{Mesinger},
  L.~{Pentericci} et~al., \emph{{The impact of scatter in the galaxy UV
  luminosity to halo mass relation on Ly {\ensuremath{\alpha}} visibility
  during the epoch of reionization}},
  \href{https://doi.org/10.1093/mnras/staa1178}{\emph{\mnras} {\bfseries 495}
  (2020) 3602} [\href{https://arxiv.org/abs/1911.03499}{{\ttfamily
  1911.03499}}].

\bibitem{Bruton2023}
S.~{Bruton}, C.~{Scarlata}, F.~{Haardt}, M.J.~{Hayes}, C.~{Mason},
  A.M.~{Morales} et~al., \emph{{The Impact of Cosmic Variance on Inferences of
  Global Neutral Fraction Derived from Ly{\ensuremath{\alpha}} Luminosity
  Functions during Reionization}},
  \href{https://doi.org/10.3847/1538-4357/acd179}{\emph{\apj} {\bfseries 953}
  (2023) 29} [\href{https://arxiv.org/abs/2305.04949}{{\ttfamily 2305.04949}}].

\bibitem{Morishita2023}
T.~{Morishita}, G.~{Roberts-Borsani}, T.~{Treu}, G.~{Brammer}, C.A.~{Mason},
  M.~{Trenti} et~al., \emph{{Early Results from GLASS-JWST. XIV. A
  Spectroscopically Confirmed Protocluster 650 Million Years after the Big
  Bang}}, \href{https://doi.org/10.3847/2041-8213/acb99e}{\emph{\apjl}
  {\bfseries 947} (2023) L24}
  [\href{https://arxiv.org/abs/2211.09097}{{\ttfamily 2211.09097}}].

\bibitem{Fan2006b}
X.~{Fan}, M.A.~{Strauss}, R.H.~{Becker}, R.L.~{White}, J.E.~{Gunn},
  G.R.~{Knapp} et~al., \emph{{Constraining the Evolution of the Ionizing
  Background and the Epoch of Reionization with
  z\raisebox{-0.5ex}\textasciitilde6 Quasars. II. A Sample of 19 Quasars}},
  \href{https://doi.org/10.1086/504836}{\emph{\aj} {\bfseries 132} (2006) 117}
  [\href{https://arxiv.org/abs/astro-ph/0512082}{{\ttfamily
  astro-ph/0512082}}].

\bibitem{McGreer2015}
I.D.~{McGreer}, A.~{Mesinger} and V.~{D'Odorico}, \emph{{Model-independent
  evidence in favour of an end to reionization by z {\ensuremath{\approx}} 6}},
  \href{https://doi.org/10.1093/mnras/stu2449}{\emph{\mnras} {\bfseries 447}
  (2015) 499} [\href{https://arxiv.org/abs/1411.5375}{{\ttfamily 1411.5375}}].

\bibitem{Schroeder2013}
J.~{Schroeder}, A.~{Mesinger} and Z.~{Haiman}, \emph{{Evidence of Gunn-Peterson
  damping wings in high-z quasar spectra: strengthening the case for incomplete
  reionization at z {\ensuremath{\sim}} 6-7}},
  \href{https://doi.org/10.1093/mnras/sts253}{\emph{\mnras} {\bfseries 428}
  (2013) 3058} [\href{https://arxiv.org/abs/1204.2838}{{\ttfamily 1204.2838}}].

\bibitem{Mortlock2011}
D.J.~{Mortlock}, S.J.~{Warren}, B.P.~{Venemans}, M.~{Patel}, P.C.~{Hewett},
  R.G.~{McMahon} et~al., \emph{{A luminous quasar at a redshift of z = 7.085}},
  \href{https://doi.org/10.1038/nature10159}{\emph{\nat} {\bfseries 474} (2011)
  616} [\href{https://arxiv.org/abs/1106.6088}{{\ttfamily 1106.6088}}].

\bibitem{Totani2014}
T.~{Totani}, K.~{Aoki}, T.~{Hattori}, G.~{Kosugi}, Y.~{Niino}, T.~{Hashimoto}
  et~al., \emph{{Probing intergalactic neutral hydrogen by the Lyman alpha red
  damping wing of gamma-ray burst 130606A afterglow spectrum at z = 5.913}},
  \href{https://doi.org/10.1093/pasj/psu032}{\emph{\pasj} {\bfseries 66} (2014)
  63} [\href{https://arxiv.org/abs/1312.3934}{{\ttfamily 1312.3934}}].

\bibitem{McQuinn2007a}
M.~{McQuinn}, A.~{Lidz}, O.~{Zahn}, S.~{Dutta}, L.~{Hernquist} and
  M.~{Zaldarriaga}, \emph{{The morphology of HII regions during reionization}},
  \href{https://doi.org/10.1111/j.1365-2966.2007.11489.x}{\emph{\mnras}
  {\bfseries 377} (2007) 1043}
  [\href{https://arxiv.org/abs/astro-ph/0610094}{{\ttfamily
  astro-ph/0610094}}].

\bibitem{McQuinn2007b}
M.~{McQuinn}, L.~{Hernquist}, M.~{Zaldarriaga} and S.~{Dutta}, \emph{{Studying
  reionization with Ly{\ensuremath{\alpha}} emitters}},
  \href{https://doi.org/10.1111/j.1365-2966.2007.12085.x}{\emph{\mnras}
  {\bfseries 381} (2007) 75} [\href{https://arxiv.org/abs/0704.2239}{{\ttfamily
  0704.2239}}].

\bibitem{Ota2008}
K.~{Ota}, M.~{Iye}, N.~{Kashikawa}, K.~{Shimasaku}, M.~{Kobayashi}, T.~{Totani}
  et~al., \emph{{Reionization and Galaxy Evolution Probed by z = 7
  Ly{\ensuremath{\alpha}} Emitters}},
  \href{https://doi.org/10.1086/529006}{\emph{\apj} {\bfseries 677} (2008) 12}
  [\href{https://arxiv.org/abs/0707.1561}{{\ttfamily 0707.1561}}].

\bibitem{Caruana2012}
J.~{Caruana}, A.J.~{Bunker}, S.M.~{Wilkins}, E.R.~{Stanway}, M.~{Lacy},
  M.J.~{Jarvis} et~al., \emph{{No evidence for Lyman {\ensuremath{\alpha}}
  emission in spectroscopy of z > 7 candidate galaxies}},
  \href{https://doi.org/10.1111/j.1365-2966.2012.21996.x}{\emph{\mnras}
  {\bfseries 427} (2012) 3055}
  [\href{https://arxiv.org/abs/1208.5987}{{\ttfamily 1208.5987}}].

\bibitem{Tilvi2014}
V.~{Tilvi}, C.~{Papovich}, S.L.~{Finkelstein}, J.~{Long}, M.~{Song},
  M.~{Dickinson} et~al., \emph{{Rapid Decline of Ly{\ensuremath{\alpha}}
  Emission toward the Reionization Era}},
  \href{https://doi.org/10.1088/0004-637X/794/1/5}{\emph{\apj} {\bfseries 794}
  (2014) 5} [\href{https://arxiv.org/abs/1405.4869}{{\ttfamily 1405.4869}}].

\bibitem{Schenker2014}
M.A.~{Schenker}, R.S.~{Ellis}, N.P.~{Konidaris} and D.P.~{Stark},
  \emph{{Line-emitting Galaxies beyond a Redshift of 7: An Improved Method for
  Estimating the Evolving Neutrality of the Intergalactic Medium}},
  \href{https://doi.org/10.1088/0004-637X/795/1/20}{\emph{\apj} {\bfseries 795}
  (2014) 20} [\href{https://arxiv.org/abs/1404.4632}{{\ttfamily 1404.4632}}].

\bibitem{Ishigaki2018}
M.~{Ishigaki}, R.~{Kawamata}, M.~{Ouchi}, M.~{Oguri}, K.~{Shimasaku} and
  Y.~{Ono}, \emph{{Full-data Results of Hubble Frontier Fields: UV Luminosity
  Functions at z {\ensuremath{\sim}} 6-10 and a Consistent Picture of Cosmic
  Reionization}}, \href{https://doi.org/10.3847/1538-4357/aaa544}{\emph{\apj}
  {\bfseries 854} (2018) 73}
  [\href{https://arxiv.org/abs/1702.04867}{{\ttfamily 1702.04867}}].

\bibitem{Furlanetto2006}
S.R.~{Furlanetto}, M.~{McQuinn} and L.~{Hernquist}, \emph{{Characteristic
  scales during reionization}},
  \href{https://doi.org/10.1111/j.1365-2966.2005.09687.x}{\emph{\mnras}
  {\bfseries 365} (2006) 115}
  [\href{https://arxiv.org/abs/astro-ph/0507524}{{\ttfamily
  astro-ph/0507524}}].

\bibitem{Madau1997}
P.~{Madau}, A.~{Meiksin} and M.J.~{Rees}, \emph{{21 Centimeter Tomography of
  the Intergalactic Medium at High Redshift}},
  \href{https://doi.org/10.1086/303549}{\emph{\apj} {\bfseries 475} (1997) 429}
  [\href{https://arxiv.org/abs/astro-ph/9608010}{{\ttfamily
  astro-ph/9608010}}].

\bibitem{Hirata2006}
C.M.~{Hirata}, \emph{{Wouthuysen-Field coupling strength and application to
  high-redshift 21-cm radiation}},
  \href{https://doi.org/10.1111/j.1365-2966.2005.09949.x}{\emph{\mnras}
  {\bfseries 367} (2006) 259}
  [\href{https://arxiv.org/abs/astro-ph/0507102}{{\ttfamily
  astro-ph/0507102}}].

\bibitem{Barkana2005}
R.~{Barkana} and A.~{Loeb}, \emph{{A Method for Separating the Physics from the
  Astrophysics of High-Redshift 21 Centimeter Fluctuations}},
  \href{https://doi.org/10.1086/430599}{\emph{\apjl} {\bfseries 624} (2005)
  L65} [\href{https://arxiv.org/abs/astro-ph/0409572}{{\ttfamily
  astro-ph/0409572}}].

\bibitem{Wouthuysen1952}
S.A.~{Wouthuysen}, \emph{{On the excitation mechanism of the 21-cm
  (radio-frequency) interstellar hydrogen emission line.}},
  \href{https://doi.org/10.1086/106661}{\emph{\aj} {\bfseries 57} (1952) 31}.

\bibitem{Field1958}
G.B.~{Field}, \emph{{Excitation of the Hydrogen 21-CM Line}},
  \href{https://doi.org/10.1109/JRPROC.1958.286741}{\emph{Proceedings of the
  IRE} {\bfseries 46} (1958) 240}.

\bibitem{Mesinger2013}
A.~{Mesinger}, A.~{Ferrara} and D.S.~{Spiegel}, \emph{{Signatures of X-rays in
  the early Universe}},
  \href{https://doi.org/10.1093/mnras/stt198}{\emph{\mnras} {\bfseries 431}
  (2013) 621} [\href{https://arxiv.org/abs/1210.7319}{{\ttfamily 1210.7319}}].

\bibitem{Barry2022}
N.~{Barry}, G.~{Bernardi}, B.~{Greig}, N.~{Kern} and F.~{Mertens},
  \emph{{SKA-low intensity mapping pathfinder updates: deeper 21 cm power
  spectrum limits from improved analysis frameworks}},
  \href{https://doi.org/10.1117/1.JATIS.8.1.011007}{\emph{Journal of
  Astronomical Telescopes, Instruments, and Systems} {\bfseries 8} (2022)
  011007} [\href{https://arxiv.org/abs/2110.06173}{{\ttfamily 2110.06173}}].

\bibitem{Abdurashidova2022a}
Z.~{Abdurashidova}, J.E.~{Aguirre}, P.~{Alexander}, Z.S.~{Ali}, Y.~{Balfour},
  A.P.~{Beardsley} et~al., \emph{{First Results from HERA Phase I: Upper Limits
  on the Epoch of Reionization 21 cm Power Spectrum}},
  \href{https://doi.org/10.3847/1538-4357/ac1c78}{\emph{\apj} {\bfseries 925}
  (2022) 221} [\href{https://arxiv.org/abs/2108.02263}{{\ttfamily
  2108.02263}}].

\bibitem{Abdurashidova2022b}
Z.~{Abdurashidova}, J.E.~{Aguirre}, P.~{Alexander}, Z.S.~{Ali}, Y.~{Balfour},
  R.~{Barkana} et~al., \emph{{HERA Phase I Limits on the Cosmic 21 cm Signal:
  Constraints on Astrophysics and Cosmology during the Epoch of Reionization}},
  \href{https://doi.org/10.3847/1538-4357/ac2ffc}{\emph{\apj} {\bfseries 924}
  (2022) 51} [\href{https://arxiv.org/abs/2108.07282}{{\ttfamily 2108.07282}}].

\bibitem{Ewall-Wice2016}
A.~{Ewall-Wice}, J.S.~{Dillon}, J.N.~{Hewitt}, A.~{Loeb}, A.~{Mesinger},
  A.R.~{Neben} et~al., \emph{{First limits on the 21 cm power spectrum during
  the Epoch of X-ray heating}},
  \href{https://doi.org/10.1093/mnras/stw1022}{\emph{\mnras} {\bfseries 460}
  (2016) 4320} [\href{https://arxiv.org/abs/1605.00016}{{\ttfamily
  1605.00016}}].

\bibitem{Barry2019}
N.~{Barry}, M.~{Wilensky}, C.M.~{Trott}, B.~{Pindor}, A.P.~{Beardsley},
  B.J.~{Hazelton} et~al., \emph{{Improving the Epoch of Reionization Power
  Spectrum Results from Murchison Widefield Array Season 1 Observations}},
  \href{https://doi.org/10.3847/1538-4357/ab40a8}{\emph{\apj} {\bfseries 884}
  (2019) 1} [\href{https://arxiv.org/abs/1909.00561}{{\ttfamily 1909.00561}}].

\bibitem{Li2019}
W.~{Li}, J.C.~{Pober}, N.~{Barry}, B.J.~{Hazelton}, M.F.~{Morales},
  C.M.~{Trott} et~al., \emph{{First Season MWA Phase II Epoch of Reionization
  Power Spectrum Results at Redshift 7}},
  \href{https://doi.org/10.3847/1538-4357/ab55e4}{\emph{\apj} {\bfseries 887}
  (2019) 141} [\href{https://arxiv.org/abs/1911.10216}{{\ttfamily
  1911.10216}}].

\bibitem{Trott2020}
C.M.~{Trott}, C.H.~{Jordan}, S.~{Midgley}, N.~{Barry}, B.~{Greig}, B.~{Pindor}
  et~al., \emph{{Deep multiredshift limits on Epoch of Reionization 21 cm power
  spectra from four seasons of Murchison Widefield Array observations}},
  \href{https://doi.org/10.1093/mnras/staa414}{\emph{\mnras} {\bfseries 493}
  (2020) 4711} [\href{https://arxiv.org/abs/2002.02575}{{\ttfamily
  2002.02575}}].

\bibitem{Yoshiura2021}
S.~{Yoshiura}, B.~{Pindor}, J.L.B.~{Line}, N.~{Barry}, C.M.~{Trott},
  A.~{Beardsley} et~al., \emph{{A new MWA limit on the 21 cm power spectrum at
  redshifts 13-17}},
  \href{https://doi.org/10.1093/mnras/stab1560}{\emph{\mnras} {\bfseries 505}
  (2021) 4775} [\href{https://arxiv.org/abs/2105.12888}{{\ttfamily
  2105.12888}}].

\bibitem{Kolopanis2019}
M.~{Kolopanis}, D.C.~{Jacobs}, C.~{Cheng}, A.R.~{Parsons}, S.A.~{Kohn},
  J.C.~{Pober} et~al., \emph{{A Simplified, Lossless Reanalysis of PAPER-64}},
  \href{https://doi.org/10.3847/1538-4357/ab3e3a}{\emph{\apj} {\bfseries 883}
  (2019) 133} [\href{https://arxiv.org/abs/1909.02085}{{\ttfamily
  1909.02085}}].

\bibitem{Patil2017}
A.H.~{Patil}, S.~{Yatawatta}, L.V.E.~{Koopmans}, A.G.~{de Bruyn},
  M.A.~{Brentjens}, S.~{Zaroubi} et~al., \emph{{Upper Limits on the 21 cm Epoch
  of Reionization Power Spectrum from One Night with LOFAR}},
  \href{https://doi.org/10.3847/1538-4357/aa63e7}{\emph{\apj} {\bfseries 838}
  (2017) 65} [\href{https://arxiv.org/abs/1702.08679}{{\ttfamily 1702.08679}}].

\bibitem{Mertens2020}
F.G.~{Mertens}, M.~{Mevius}, L.V.E.~{Koopmans}, A.R.~{Offringa}, G.~{Mellema},
  S.~{Zaroubi} et~al., \emph{{Improved upper limits on the 21 cm signal power
  spectrum of neutral hydrogen at z {\ensuremath{\approx}} 9.1 from LOFAR}},
  \href{https://doi.org/10.1093/mnras/staa327}{\emph{\mnras} {\bfseries 493}
  (2020) 1662} [\href{https://arxiv.org/abs/2002.07196}{{\ttfamily
  2002.07196}}].

\bibitem{Gehlot2020}
B.K.~{Gehlot}, F.G.~{Mertens}, L.V.E.~{Koopmans}, A.R.~{Offringa},
  A.~{Shulevski}, M.~{Mevius} et~al., \emph{{The AARTFAAC Cosmic Explorer:
  observations of the 21-cm power spectrum in the EDGES absorption trough}},
  \href{https://doi.org/10.1093/mnras/staa3093}{\emph{\mnras} {\bfseries 499}
  (2020) 4158} [\href{https://arxiv.org/abs/2010.02269}{{\ttfamily
  2010.02269}}].

\bibitem{Paciga2013}
G.~{Paciga}, J.G.~{Albert}, K.~{Bandura}, T.-C.~{Chang}, Y.~{Gupta},
  C.~{Hirata} et~al., \emph{{A simulation-calibrated limit on the H I power
  spectrum from the GMRT Epoch of Reionization experiment}},
  \href{https://doi.org/10.1093/mnras/stt753}{\emph{\mnras} {\bfseries 433}
  (2013) 639} [\href{https://arxiv.org/abs/1301.5906}{{\ttfamily 1301.5906}}].

\bibitem{HERA2022}
{The HERA Collaboration}, Z.~{Abdurashidova}, T.~{Adams}, J.E.~{Aguirre},
  P.~{Alexander}, Z.S.~{Ali} et~al., \emph{{Improved Constraints on the 21 cm
  EoR Power Spectrum and the X-Ray Heating of the IGM with HERA Phase I
  Observations}}, \href{https://doi.org/10.48550/arXiv.2210.04912}{\emph{arXiv
  e-prints} (2022) arXiv:2210.04912}
  [\href{https://arxiv.org/abs/2210.04912}{{\ttfamily 2210.04912}}].

\bibitem{Eastwood2019}
M.W.~{Eastwood}, M.M.~{Anderson}, R.M.~{Monroe}, G.~{Hallinan}, M.~{Catha},
  J.~{Dowell} et~al., \emph{{The 21 cm Power Spectrum from the Cosmic Dawn:
  First Results from the OVRO-LWA}},
  \href{https://doi.org/10.3847/1538-3881/ab2629}{\emph{\aj} {\bfseries 158}
  (2019) 84} [\href{https://arxiv.org/abs/1906.08943}{{\ttfamily 1906.08943}}].

\bibitem{Schechter1976}
P.~{Schechter}, \emph{{An analytic expression for the luminosity function for
  galaxies.}}, \href{https://doi.org/10.1086/154079}{\emph{\apj} {\bfseries
  203} (1976) 297}.

\bibitem{Sheth1999}
R.K.~{Sheth} and G.~{Tormen}, \emph{{Large-scale bias and the peak background
  split}},
  \href{https://doi.org/10.1046/j.1365-8711.1999.02692.x}{\emph{\mnras}
  {\bfseries 308} (1999) 119}
  [\href{https://arxiv.org/abs/astro-ph/9901122}{{\ttfamily
  astro-ph/9901122}}].

\bibitem{Jenkins2001}
A.~{Jenkins}, C.S.~{Frenk}, S.D.M.~{White}, J.M.~{Colberg}, S.~{Cole},
  A.E.~{Evrard} et~al., \emph{{The mass function of dark matter haloes}},
  \href{https://doi.org/10.1046/j.1365-8711.2001.04029.x}{\emph{\mnras}
  {\bfseries 321} (2001) 372}
  [\href{https://arxiv.org/abs/astro-ph/0005260}{{\ttfamily
  astro-ph/0005260}}].

\bibitem{Bouwens2015}
R.J.~{Bouwens}, G.D.~{Illingworth}, P.A.~{Oesch}, M.~{Trenti}, I.~{Labb{\'e}},
  L.~{Bradley} et~al., \emph{{UV Luminosity Functions at Redshifts z
  {\ensuremath{\sim}} 4 to z {\ensuremath{\sim}} 10: 10,000 Galaxies from HST
  Legacy Fields}},
  \href{https://doi.org/10.1088/0004-637X/803/1/34}{\emph{\apj} {\bfseries 803}
  (2015) 34} [\href{https://arxiv.org/abs/1403.4295}{{\ttfamily 1403.4295}}].

\bibitem{Koopmans2015}
L.~{Koopmans}, J.~{Pritchard}, G.~{Mellema}, J.~{Aguirre}, K.~{Ahn},
  R.~{Barkana} et~al., \emph{{The Cosmic Dawn and Epoch of Reionisation with
  SKA}},  in \emph{Advancing Astrophysics with the Square Kilometre Array
  (AASKA14)}, p.~1, Apr., 2015, \href{https://doi.org/10.22323/1.215.0001}{DOI}
  [\href{https://arxiv.org/abs/1505.07568}{{\ttfamily 1505.07568}}].

\bibitem{Mellema2015}
G.~{Mellema}, L.~{Koopmans}, H.~{Shukla}, K.K.~{Datta}, A.~{Mesinger} and
  S.~{Majumdar}, \emph{{HI tomographic imaging of the Cosmic Dawn and Epoch of
  Reionization with SKA}},  in \emph{Advancing Astrophysics with the Square
  Kilometre Array (AASKA14)}, p.~10, Apr., 2015,
  \href{https://doi.org/10.22323/1.215.0010}{DOI}
  [\href{https://arxiv.org/abs/1501.04203}{{\ttfamily 1501.04203}}].

\end{thebibliography}\endgroup







\end{document}